\documentclass[journal]{IEEEtran}

\usepackage[T1]{fontenc}
\usepackage{textcomp}

\usepackage{amsmath,amsfonts,amssymb}

\usepackage{newtxtext,newtxmath}

\usepackage{graphicx}
\usepackage[caption=false,font=normalsize,labelfont=sf,textfont=sf]{subfig}

\usepackage{subfig}
\usepackage{algorithm}
\usepackage{algpseudocode}
\usepackage{multirow}
\usepackage{array}
\usepackage{tikz}
\usetikzlibrary{shapes,arrows}
\usepackage{cite}
\usepackage{url}
\usepackage{verbatim}

\begin{document}

\title{Real-time identification and control of influential pandemic regions using graph signal variation}

\author{Sudeepini Darapu, Subrata Ghosh, Dibakar Ghosh, Chittaranjan Hens, Santosh Nannuru
\thanks{S.Darapu. is funded by a research fellowship from IIIT Hyderabad I-HUB. C.Hens. is supported by a DST-INSPIRE Faculty Award (Grant No. IFA17-PH193). S.Nannuru. is also supported by a DST-INSPIRE Faculty Award (Grant No. IFA17-ENG236).}
\thanks{Manuscript received ?? revised ??.}}



\maketitle

\begin{abstract}
The global spread of pandemics is facilitated by the mobility of populations, transforming localized infections into widespread phenomena. To contain it, timely identification of influential regions that accelerate this process is necessary. In this work, we model infection as a temporally evolving graph signal and propose graph signal variation-based metrics to capture spatio-temporal changes. Both graph domain and time domain locality are modeled. Based on this metric, we propose an online algorithm to identify influential regions. Simulations demonstrate that the proposed method effectively identifies geographical regions with a higher capacity to spread the infection. Isolating these regions leads to a significant reduction in cumulative infection. Simulations, along with analyses of hybrid H1N1 data and real-world Indian COVID-19 data, underscore the utility of proposed metric in enhancing our understanding and control of infection spread.
\end{abstract}


\section{Introduction}
\label{sec:intro}

\IEEEPARstart{O}ver the past decades, the world has faced various \cite{baker2022infectious} health challenges --  SARS outbreak in $2004$ \cite{colizza2007predictability}, H$1$N$1$ in $2009$ \cite{novel2009emergence}, and recently COVID-$19$ \cite{arenas2020modeling,kraemer2020effect}. The COVID-$19$ pandemic has demonstrated how infection can spread rapidly, especially in large gatherings that lack precautions and unrestricted travel to highly infected or rapidly transmitting zones. These events emphasize the need for effective strategies to control the spread of infection.

This study aims to identify \emph{influential regions} -- regions that play a substantial role in infection propagation. Specifically, we study a network of interconnected geographical regions that allow the movement of individuals across them. The interconnections facilitate disease transmission via human migration. Depending on the modeling framework, the size of the geographical \emph{region} varies. For example, in one of our analyses, the regions are districts and the network spans over the country, whereas in a different analysis, regions model airports while the network is spread across multiple countries. Isolating influential regions is essential for understanding and effectively controlling the disease spread within a network.

Often, the influential regions have high disease prevalence, but not always. For example, a region could be influential due to its strategic position in the network \emph{or} due to the infection dynamics in its neighbourhood \emph{or} both. Thus, we use the network structure \emph{as well as} the network infection dynamics to identify the influential regions.

Such a syncretic study of signals and networks forms the basis of graph signal processing (GSP) \cite{shuman2013emerging, sandryhaila2014discrete, sandryhaila2013discrete, ortega2018graph}. 
For time-series signals in the network domain, the GSP framework allows us to account for both relational and temporal information. We borrow tools from GSP to propose novel variation metrics for temporally evolving graph signals. Subsequently, we rank the network nodes according to their capacity to ``influence'', i.e., spread the infection.

\subsection{Related work}

Most real-world influential node identification methods have ad hoc design. While factors like confirmed cases, infection rates, and local healthcare capacity are implied considerations, explicit documentation is lacking. The focus is on regional disease parameters and lacks comprehensive analysis. Some work has been done on the analysis of disease spread on epidemiological networks using graph theory \cite{sabidussi1966centrality, chen2012identifying, freeman2002centrality, bonacich2007some, brin1998pagerank, vynnycky2010introduction, dietz1993estimation, ji2023signal}  and graph signal processing \cite{zhao2015identifying, geng2022analysis, pena2016source, li2021graph}. While graph theory focuses solely on the network topology, graph signal processing utilizes both the topology and the signal.

Networks have a wide diversity of nodes and the node characteristics play a crucial role in spread of the disease in epidemic networks. 
To capture this, various centrality measures have been devised which rank the nodes in order of their importance or influence in the network. Depending on the measure used, they are called closeness centrality \cite{liu2018identifying,sabidussi1966centrality}, degree centrality, clustering coefficient \cite{song2017clustering, chen2012identifying}, betweenness centrality \cite{wei2022identifying, freeman2002centrality}, eigenvector centrality \cite{huang2021identifying, brin1998pagerank, bonacich2007some} and $k$-shell decomposition~\cite{kitsak2010identification}. These graph theory based methods focus only on network connections, ignoring the spatio-temporal spread of infection. 

Integrating signals with the graph structure enhances influential node identification by capturing crucial local and temporal dynamics. Recently, methods have been proposed to study the disease spread using tools from GSP. In \cite{zhao2015identifying}, graph signal processing based centrality is proposed for identifying influential nodes, considering both node heterogeneity and network topology. Graph frequency analysis has been used \cite{li2021graph} to study spread of COVID-19 in US counties. In \cite{pena2016source}, authors process ``John Snow’s Cholera Data'' and model cholera transmission as heat diffusion process to localize the source of infection. Using the spectral graph wavelet transform, spatio-temporal patterns of COVID-19 virus spread in Massachusetts are analysed in \cite{geng2022analysis}. 
Notions of smoothness and generalized smoothness are defined in \cite{hosseinalipour2017detection} for infection detection but it does not account for temporal variability.

Spatio-temporal studies in \cite{zhu2022early,purwanto2021spatiotemporal,ghosh2021reservoir,bao2022impact} perform identification of COVID hotspots (influential regions). In \cite{zhu2022early} a spatio-temporal Bayesian framework is implemented for early detection of COVID-$19$ hotspots using deep neural networks. However, neural networks require extensive training and are computationally demanding. The study in \cite{purwanto2021spatiotemporal} utilizes two spatio-temporal analysis models, namely the emerging hotspot analysis and space–time cubes, to identify temporal hotspot clusters associated with COVID-$19$. A straightforward method to identify influential regions would be to pick the regions with maximum infection.
However, it does not consider how infections spread through the network, which limits its effectiveness.

\subsection{Contributions}

To address various drawbacks of existing methods, we propose a real-time method for influential region identification that uses the network connections as well as the spatio-temporal infection dynamics, unified within the GSP framework. Our main contributions are:
\begin{enumerate}
    \item A novel metric -- temporal local variation (TLV) -- is introduced for temporally varying graph signals.
    \item A graph signal variation-based online algorithm is developed to identify influential nodes.
    \item The effectiveness of the proposed methods are validated across various topology and infection propagation scenarios.
    \item A control measure study is conducted to quantitatively compare the proposed methods with the literature.
    \item The proposed methods are applied to synthetically generated H1N1 infection data derived from SIR dynamics on the global airport network, as well as real-world COVID-19 infection data from districts across India.
\end{enumerate}

The paper is organized as follows: graph signal preliminaries are discussed in Section \ref{sec:GSV}. Signal variation metrics for temporally evolving graph signals are proposed in Section \ref{sec:mainTLV}. Methods for network construction and SIR model for epidemic dynamics are detailed in Section \ref{sec:NED}. Algorithms for identifying influential nodes and implementing the control strategy are described in Section \ref{sec:ICIR}. Simulation results including influential node identification  and the effects of control measures, are presented in Section \ref{sec:SAR}. Section \ref{sec:realdata} focuses on analyzing real-world data from the COVID-19 infection cases in India, including a comparison with few identified super spreader events in the country \cite{althouse2020superspreading, kumar2020significance}.

In accordance with established convention, matrices are consistently denoted by bold uppercase letters, vectors are expressed using bold lowercase letters, and scalars are represented by non-bold uppercase and lowercase letters.

\section{Graph signal preliminaries}
\label{sec:GSV}
Let $\mathcal{G} = (\mathcal{V}, \mathcal{E}, \mathbf{W})$ represent an undirected weighted graph, where $\mathcal{V} = \{1, 2, \ldots, N\}$ is the set of nodes, $\mathcal{E}$ is the edge set, and $\mathbf{W}$ is the weighted adjacency matrix. The weight $w_{ij}$ signifies the non-negative edge weight between nodes $i$ and $j$. Degree $d_i$ of the $i$th node is the sum of weights of all edges connected to node $i$. The graph Laplacian is defined as $\mathbf{L} = \mathbf{D} - \mathbf{W}$, where $\mathbf{D}$ is the diagonal degree matrix. Let $\mathbf{x}_t = [x_t(1), x_t(2), \ldots, x_t(N)]^T$ represent the graph signal at time $t$ and consecutive $T$ such graph signals are compactly arranged in the matrix
$\mathbf{X} = [\mathbf{x}_1, \mathbf{x}_2, \ldots, \mathbf{x}_T]$. 

The graph Fourier transform (GFT) \cite{sandryhaila2013discrete} analyzes the spectral characteristics of graph signal. It provides an alternate representation of the graph signals using vectors in the eigenspace of the graph Laplacian $\mathbf{L}$. Let $\mathbf{L} = \mathbf{U}\mathbf{\Lambda}\mathbf{U}^{-1}$ be the eigenvalue decomposition. The eigenvalue matrix $\mathbf{\Lambda} = \textit{diag}(\lambda_1, \ldots , \lambda_N)$ contains ordered non-negative eigenvalues ($\lambda_N > \lambda_{N-1} \ldots \lambda_1 \geq 0$) and the eigenvector matrix $\mathbf{U} = [\mathbf{u}_1, \ldots, \mathbf{u}_N]$ contains orthonormal eigenvectors. The GFT of a graph signal $\mathbf{x}_t$ is,
\begin{equation}\label{eq:GFT}
    \Tilde{\mathbf{x}}_t = \mathbf{U}^{-1} \mathbf{x}_t.
\end{equation}
Eigenvectors corresponding to large eigenvalues capture the rapid variations in the graph signal and vice versa. Thus eigenvalues play the role of frequency \cite{sandryhaila2013discrete}.

\subsection{Graph high pass filter ($\text{HPF}_{\mathcal{G}}$)}
The graph high pass filter \cite{sandryhaila2013discreteF} nullifies the influence of low-frequency components in the graph signal. Let $\mathbf{h}_{\text{HPF}}$ be an $N$-length binary vector indicating positions of highest $l < N$ graph frequencies. High pass filtering operation is given by,
\begin{align}
    \mathbf{X}_{\text{HPF}} &= \mathbf{H}_{\text{HPF}} \mathbf{X} \,, \label{eq:HPF} \\
    \mathbf{H}_{\text{HPF}} &= \mathbf{U} \mathbf{\Tilde{H}}_{\text{HPF}} \mathbf{U}^T \,, \quad 
    \mathbf{\Tilde{H}}_{\text{HPF}} = \textit{diag}(\mathbf{h}_{\text{HPF}}) \,.
\end{align}
We use a $\text{HPF}_{\mathcal{G}}$ corresponding to the highest $25\%$ of eigenvalues. The filtered signal ($\mathbf{X}_{\text{HPF}}$) is used to identify the top $p\%$ of nodes as influential nodes, as outlined in Algorithm \ref{Alg:Online}.

\subsection{Spectral graph wavelet transform (SGWT)}
\label{sec:SGWT}

The SGWT \cite{hammond2011wavelets} is used for analyzing graph signals to uncover localized details and used as multi-resolution analysis tool \cite{Akansu1992}. Graph wavelets are specialized functions within the graph's spectral domain. These tailored wavelets systematically unveil smooth and oscillatory patterns. In this study, we employ SGWT to identify influential nodes. We implemented methodology analogous to that proposed in \cite{geng2022analysis} for SGWT processing.

\section{Variation of temporally evolving graph signals}
\label{sec:mainTLV}
Total variation \cite{sandryhaila2015variation} is an important measure of a graph signal which captures variability of the signal with respect to the underlying graph. Though useful, it falls short of fully capturing variability of temporally evolving graph signals, since the underlying domain now includes both graph and time. Additionally, total variation computes a single scalar metric for the complete graph, and is thus unable to capture local signal dynamics (at the node level). To address these limitations, we introduce the following signal variation measures -- local variation (LV) and temporal local variation (TLV).

\subsection{Total variation and local variation (LV)} 
\label{sec:TVLV}

The total variation \cite{sandryhaila2015variation} of a graph signal $\mathbf{x}_{t}$ at time $t$ is,
\begin{equation}
{TV(t)} = \sum_{i=1}^{N} \sum_{j\in \mathcal{N}_i}{{(x_{t}(i) - x_{t}(j))}^2 w_{ij}}=\mathbf{x_{t}}^T \mathbf{L} \mathbf{x_{t}} \,,
\label{eq2}
\end{equation}
where $\mathcal{N}_i$ denotes the set of neighbors of node $i$. A graph signal's total variation indicates how much it changes globally over $\mathcal{G}$. In order to measure how much a graph signal varies locally, we define the local variation at node $i$ as,
\begin{equation}\label{eq:LV}
LV(i,t) = \sum_{j\in \mathcal{N}_i}{{(x_{t}(i)-x_{t}(j))}^2 w_{ij}}= \mathbf{X}_{LV}.
\end{equation}
Thus local variation is a measure of variation with focus on node $i$. Large $LV(i,t)$ values imply that the signal at node $i$ is highly varying compared to its neighbors, whereas small $LV(i,t)$ implies the signal at node $i$ does not vary much compared to its neighbors. Taken together for all the nodes, at time $t$, local variation is an $N$-length vector. By definition, sum of the local variation over all the nodes equals total variation. Both total variation and local variation are non-negative.


\subsection{Temporal local variation (TLV)}
\label{sec:TLV}

The local variation defined above captures nodal signal dynamics with respect to its neighbors, but is unaware of temporal dynamics. This is crucial for temporally evolving graph signals. Hence, we extend local variation to encompass temporal dynamics and define temporal local variation (TLV) at time $t$ and node $i$ as,
\begin{align}
TLV(i,t) 
&= \alpha \, \phi(i,t) \left( {x}_{t}(i) - {x}_{t-1}(i) \right)^2 \nonumber \\ 
& \;\; + (1-\alpha) \sum_{j\in \mathcal{N}_i} \left( {x}_{t}(i) - {x}_{t}(j) \right)^2 w_{ij} \,, \\
\label{eq:TLV}
&= \alpha \, \phi(i,t) \, \text{Temporal variation} \nonumber \\
& \quad + (1-\alpha) \, \text{Local variation} \,,
\end{align}
where, $\phi(i,t)=\text{sign}(({x}_{t}(i)-{x}_{t-1}(i)))$ captures directional information of temporal change and the weight factor $\alpha \in [0,1]$ determines the degree of emphasis placed on temporal and local (neighbourhood) aspects of variations. When $\alpha = 0$, it disregards temporal gradient and when $\alpha = 1$, the local neighborhood gradient is ignored.
When compared to LV, TLV offers a more holistic view of signal dynamics as it captures both temporal (TV) and local variations. Both magnitude and direction of temporal changes are captured by TLV. The parameter $\alpha$ allows precise adjustment of temporal and local variation balance based on the application.

Often, we are interested in comparing the signal variation across nodes and time. To make this task easier, we perform normalization. Local variation is normalized by dividing by the maximum (across nodes) value of local variation.
For a windowed data $\mathbf{X}_{N\times r}$ of $r$ time steps, temporal variation ($\mathbf{X}_{TV}$) and local variation ($ \mathbf{X}_{LV}$) are computed using \eqref{eq:LV} and \eqref{eq:TV}.  Subsequently, they are normalized by their maximum value in the windowed data. Scaled variations are combined to compute the normalized TLV as in \eqref{eq:TLVN}. 

\begin{align}
TV(i,t)  = \mathbf{X}_{TV}=
\begin{cases} 
    0, & t = 1 \\
    ({x}_{t}(i) - {x}_{t-1}(i))^2, & t > 1
\end{cases}
\label{eq:TV}
\end{align}

\begin{align}
    TLV_N(i,t) &= \alpha \frac{\phi(i,t) \left( {x}_{t}(i) - {x}_{t-1}(i) \right)^2}{\text{max}(\mathbf{X}_{TV})} \nonumber \\
    & \;\; + (1-\alpha) \frac{\sum_{j \in \mathcal{N}_i} \left( {x}_{t}(i) - {x}_{t}(j) \right)^2 w_{ij} }{\text{max}(\mathbf{X}_{LV})}\,,
     \label{eq:TLVN}
\end{align}
This normalization ensures that local variation ranges from $0$ to $1$, while temporal variation ranges from $-1$ to $1$. Consequently, TLV ranges from $-1$ to $2$. A numerical illustration of TLV along with an analysis of $\alpha$, is provided in the supplementary material \cite{SDP}.

\section{Networks and epidemic dynamics}
\label{sec:NED}
In this section, we discuss synthetic topology construction and epidemic infection data generation. Our modeling incorporates Susceptible-Infected-Recovered (SIR) dynamics of interaction and diffusion \cite{colizza2007predictability, keeling2008modeling, brockmann2013hidden, hens2019spatiotemporal, belik2011natural}, where individuals interact within local groups and swiftly diffuse (migrate) between regions. Additionally, generation of H1N1 data using a real-world network and epidemic model is discussed.

\subsection{Network construction}
\label{sec:NetworkConstruction}
Two heterogeneous topological networks with distinct characteristics are generated. 

\subsubsection{Random distance-based graph}
\begin{figure*}[t]
    \centering
    \includegraphics[width=1\linewidth]{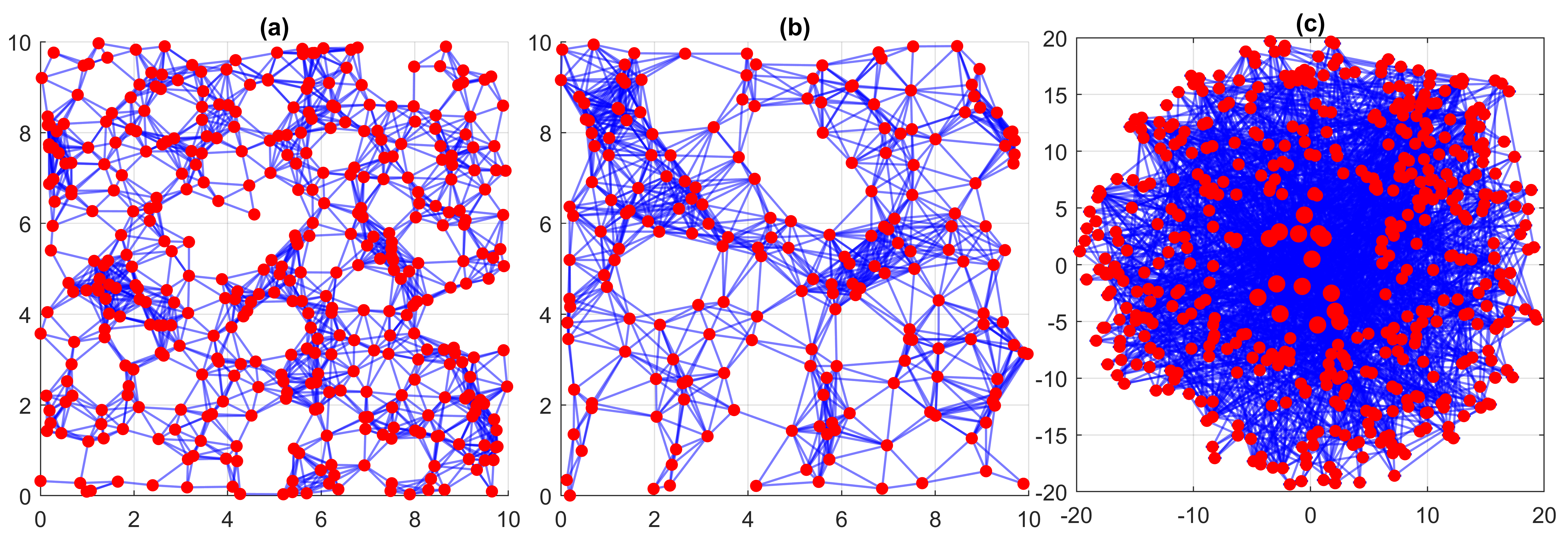} 
    \caption{(a) Random distance-based graph, $N=400$, (b) Random distance-based graph, $N=200$, (c) Scale free graph, $N=500$}
    \label{fig:graphs}
\end{figure*}

The random distance-based graph represents a network where connections between nodes are established based on their proximity within a specified distance threshold. This type of graph is characterized by short-range connections. We perform uniform sampling of points within a square of dimension $D$ to yield location attributes for the $N$ graph nodes. The weights between nodes are computed as, 
\begin{align}\label{Weights}
    W_{ij} =
        \begin{cases}
            e^{\frac{-{\text{d}(i,j)}^2}{\sigma^2}}, \; \text{d}(i,j) \leq \vartheta, & i \neq j \\
            0, & i=j,
        \end{cases}
\end{align}
where $\text{d}(i,j)$ is the Euclidean distance between nodes $i$ and $j$ and ${\sigma}^2$ a scaling parameter. Figures \ref{fig:graphs}a and \ref{fig:graphs}b depict simulated random distance-based graphs. They have diverse connection densities similar to real-world road transport networks, where urban areas often exhibit denser connectivity in comparison to the sparser connections observed in rural areas. 

\subsubsection{Scale free graph}
The scale-free network \cite{barabasi1999emergence} is a type of complex network with a power-law degree distribution -- a small number of nodes with very high degrees (hubs) and a large number of nodes with relatively low degrees. This distribution is frequently encountered in various real-world systems, including social networks, the World Wide Web, and air transportation networks featuring a mix of both short and long-range connections. Figure \ref{fig:graphs}c depicts a scale free graph. Hubs, represented by larger circles, are centrally positioned, while smaller circles signify low degree nodes, dispersed around these hubs. In this study, we use binary scale-free graph.

\subsection{Epidemic dynamics}
\label{sec:EpidemicDynamics}
Two epidemic models are discussed and implemented: the network SIR model and the H1N1 data generation model for simulating disease spread in interconnected populations.

\subsubsection{SIR dynamics}
\label{sec:SIRDynamics}
\begin{flushleft}
We consider the standard SIR (susceptible-infected-recovered) model without vital dynamics such as birth and death details to depict how an epidemic propagates through a population. In SIR dynamics \cite{keeling2008modeling}, a set of $3$-coupled equations describe how the disease spreads within a location,
\end{flushleft}

\begin{eqnarray}
\label{eq:model} 
\dot{S}(t) &=& \displaystyle  -\beta \frac{S(t) I(t)}{H} ,\\
\dot{I}(t)  &=& \displaystyle \beta \frac{S(t) I(t)}{H} - \gamma I(t), \\   
\dot{R}(t) &=&  \gamma I(t),
\end{eqnarray}
where susceptible ($S$), infected ($I$), and recovered ($R$) are the three ``compartments'' of the population. The rates of disease transmission and average recovery are $\beta$ and $\gamma$, respectively. $H$ is the population size of that location. We now consider a network of $N$ locations and employ a coupled population (meta-population) network with SIR dynamics. Let the $i^{th}$ location contain $H_i (i=1,2,...,N)$ individuals. We formulate the meta-population network equations  \cite{brockmann2013hidden,ghosh2021optimal} by taking into account the dispersion through the diffusion of normalized susceptible ($\mathcal{S}_i=\frac{S_i}{H_i}$), infected ($\mathcal{I}_i=\frac{I_i}{H_i}$), and recovered ($\mathcal{R}_i=\frac{R_i}{H_i}$) individuals,
   \begin{eqnarray}
   	\begin{array}{llll}
   		\dot{\mathcal{S}}_i(t) &=& \displaystyle -\beta_{i}  \mathcal{S}_i \mathcal{I}_i  + \frac {\kappa}{d_{i}} \sum_{j\in \mathcal{N}_i} W_{ij} (\mathcal{S}_{j}-\mathcal{S}_{i}), \\
   		\dot{\mathcal{I}}_i(t) &=& \displaystyle \beta_{i}  \mathcal{S}_i \mathcal{I}_i -\displaystyle \gamma_{i} \mathcal{I}_i +
   		\frac {\kappa}{d_{i}} \sum_{j\in \mathcal{N}_i} W_{ij} (\mathcal{I}_{j}-\mathcal{I}_{i}).\\
            \dot{\mathcal{R}}_i(t) &=& \displaystyle \gamma_{i} \mathcal{I}_i +
   		\frac {\kappa}{d_{i}} \sum_{j\in \mathcal{N}_i} W_{ij} (\mathcal{R}_{j}-\mathcal{R}_{i})
   	\end{array}
   	\label{eq:network_model}   
   \end{eqnarray}
Here $\mathbf{W}$ is the weighted adjacency matrix representing the local network and $d_i=\sum_{j\in \mathcal{N}_i} W_{ij}$ is the degree of the $i^{\rm th}$ node. The population migration is modeled through the diffusive term $\sum_{j\in \mathcal{N}_i} w_{ij} (Y_{j}-Y_{i})$ connected through two compartments ${Y: \mathcal{S}_i,\mathcal{I}_i,\mathcal{R}_i}$. Strength of the migration is determined by $\kappa$. To solve the ODE equations, we utilised Runge-Kutta (RK4) \cite{runge1895numerische} method with adaptive step size. 

In our study, we make the assumption that, when generating data using the SIR model, the disease transmission rate ($\beta$) is same across all the nodes. In real-world scenarios, this may vary across nodes and time. This simplification allows us to establish a baseline framework for analysis, acknowledging the need for further exploration and refinement when incorporating real-world complexities into the model. The infection fraction $\mathcal{I}_i$ is used as the graph signal in our analysis. Henceforth, we use the term `infection' interchangeably with `infection fraction'.

\subsubsection{Hybrid model -- H1N1 data}
In order to confirm the dependability of our approach and demonstrate its practicality in real-world pandemic situations, we employed a hybrid model that combines the dynamics of H1N1 with the global airport network as its foundation. We use the global airport transportation network \cite{hens2019spatiotemporal, brockmann2013hidden,  brockmann2013hiddenSupple} comprising of 1292 airports (representing nodes) and integrate SIR network dynamics with disease-specific parameters related to H1N1 to mimic the H1N1 pandemic, as described in~\cite{brockmann2013hidden}. The diffusive coupling employed in our model represents the movement of the air passengers across the network. All the parameters are taken from~\cite{brockmann2013hidden}.

\section{Identification and control of influential regions}
\label{sec:ICIR}

\subsection{Online identification of influential nodes}

Given a graph $\mathcal{G}$ and the graph signal $\mathbf{X}$, our goal is to identify the nodes that significantly influence the evolution of the graph signal over time. We present a methodology for online identification of influential nodes using graph signal variations. Although we exclusively study epidemiological networks, our approach could potentially be used to process other multivariate time-series data with an associated graph.

The methodology is presented in Algorithm \ref{Alg:Online}. It takes two inputs: the graph $\mathcal{G} = (\mathcal{V}, \mathcal{E}, \mathbf{W})$ and the graph signal $\mathbf{X}$. Data within a temporal window of length $r$, i.e. from $t-r+1$ to $t$, is processed at time $t$. For this data $\overline{\mathbf{X}}$, the signal variation $\overline{\mathbf{X}}_{var}$ is computed using any of the three different techniques -- $\text{HPF}_{\mathcal{G}}$, LV, and TLV. The last column $\overline{\mathbf{X}}_{var}(:,r)$ represents the variation at time $t$. After sorting it to obtain $\overline{\mathbf{x}}_{\text{sort}}$, we identify the top $p\%$ nodes with the largest graph signal variation and label them as influential nodes at time step $t$. For comparison, we implemented a baseline approach in which the sorted infection of each node is used to identify the top $p\%$ nodes. We call this approach as ``Max'' processing.

\begin{algorithm} 
	\caption{Online identification of influential nodes}
	\begin{algorithmic}[1]
		\For {$t=r,r+1,2,\ldots$}
				\State $\overline{\mathbf{X}}=\mathbf{X}(:,t-r+1:t)$
				\State 
                Compute $ \overline{\mathbf{X}}_{var}$ using \eqref{eq:HPF} or \eqref{eq:LV} or \eqref{eq:TLV} or SGWT
                \State $\overline{\mathbf{x}}_{sort}=\text{sort}(\overline{\mathbf{X}}_{var}(:,r), descend)$
                \State top $p\%$ nodes of  $\overline{\mathbf{x}}_{sort}$ are influential nodes at time step $t$
			\EndFor
	\end{algorithmic}
 \label{Alg:Online}
\end{algorithm}

\subsection{Control measures}
We study the effect of isolating influential regions from rest of the network in containing the infection. This infection control measure provides a quantitative way to compare influential regions identified by various methods. Algorithm \ref{Alg:ControlAlgorithm} outlines our approach for reducing disease propagation. Let $m$ influential nodes be identified at time $T_c$. At time $T_c+1$, we remove all the neighbor connections of influential nodes to generate control graph $\mathcal{G}_c$. It is used to minimize the spread of infection across the network. This new graph $\mathcal{G}_c$ is used for disease propagation \eqref{eq:network_model} from $T_c+1$ along with infection and susceptible data at time $T_c$ as the initial infection state.

\begin{algorithm}
\caption{Influential node-based control measure}
\begin{algorithmic}[1]
\Require $\mathcal{G}$, time step $T_c$, $\mathbf{X}$
\State Identify influential nodes: Given $\mathcal{G}$ and $\mathbf{X}$, identify $m$ influential nodes at $T_c$ using Algorithm \ref{Alg:Online}.
\State Generate control graph: At $T_c+1$, generate $\mathcal{G}_c$ by disconnecting neighbors of influential nodes
\State Disease propagation: Use $\mathcal{G}_c$ and infection at $T_c$ to
 simulate propagation \eqref{eq:network_model} from $T_c+1$.
\end{algorithmic}
\label{Alg:ControlAlgorithm}
\end{algorithm}

\section{Simulations and results}
\label{sec:SAR}

\subsection{Epidemic data generation}
\label{Sec:Epidemic data generation}
In order to validate our methods, we use two heterogeneous networks. Figure \ref{fig:graphs}a and \ref{fig:graphs}b depict random distance-based graphs with $N=400 \, (\vartheta=1.95)$ and $N=200 \, (\vartheta=1.5)$ generated using $D=10$. A scale free graph of size $N = 500$ is illustrated in Figure \ref{fig:graphs}c with average degree of $6$ and scaling exponent $3$. We consider three scenarios for epidemic data generation using SIR dynamics:
\begin{enumerate}
    \item \emph{Single perturbation:} The epidemic is initiated by selecting a single source node (randomly chosen) and infecting a small fraction ($0.2\%$) of the total population. Constant transmission rates ($\beta$) and recovery rates ($\gamma$) are maintained for all network nodes. At the initial time step, all nodes, except the source node, are infection-free with a relatively high coupling coefficient ($\kappa$).
    \item \emph{Double perturbation:} This is similar to the previous scenario, except we introduce a secondary perturbation to the same source node by increasing the infection rate at the time step when the first perturbed signal reaches its peak value. This perturbation involves a temporal adjustment of the $\beta$, transitioning from its initial value to a modified value at a specific time point, while $\kappa$ remains constant.
    \item \emph{Multiple infections:} In this scenario, we generate data by initially introducing infections at a few randomly selected source nodes. At subsequent time steps, the infection rate is increased for different sets of randomly chosen nodes, ensuring no overlap with the previously infected nodes. This process is repeated across multiple time steps while $\kappa$ remains constant.
\end{enumerate}
For these scenarios, more details and simulation parameter values are discussed in Section \ref{sec:analysis_spread}.

\subsection{Data visualization}
\begin{figure*}[t]
    \centering
    \includegraphics[width=0.21\linewidth]{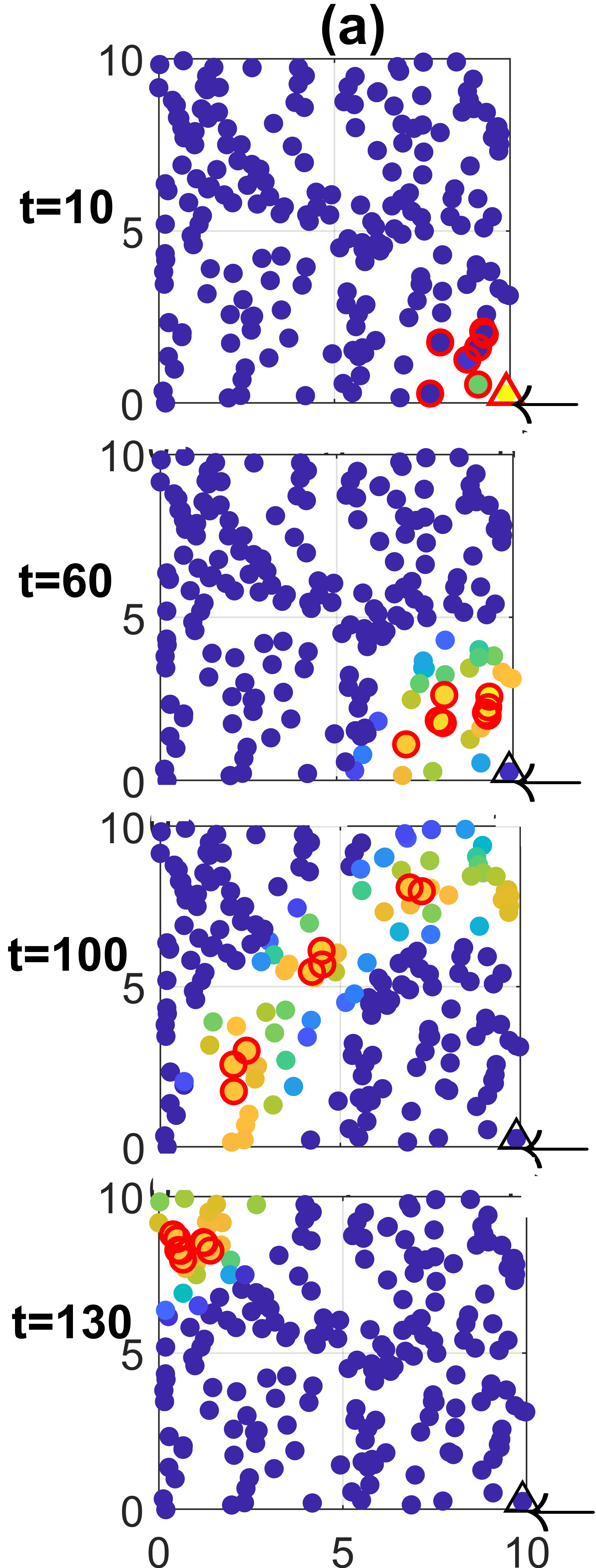} 
    \includegraphics[width=0.20\linewidth]{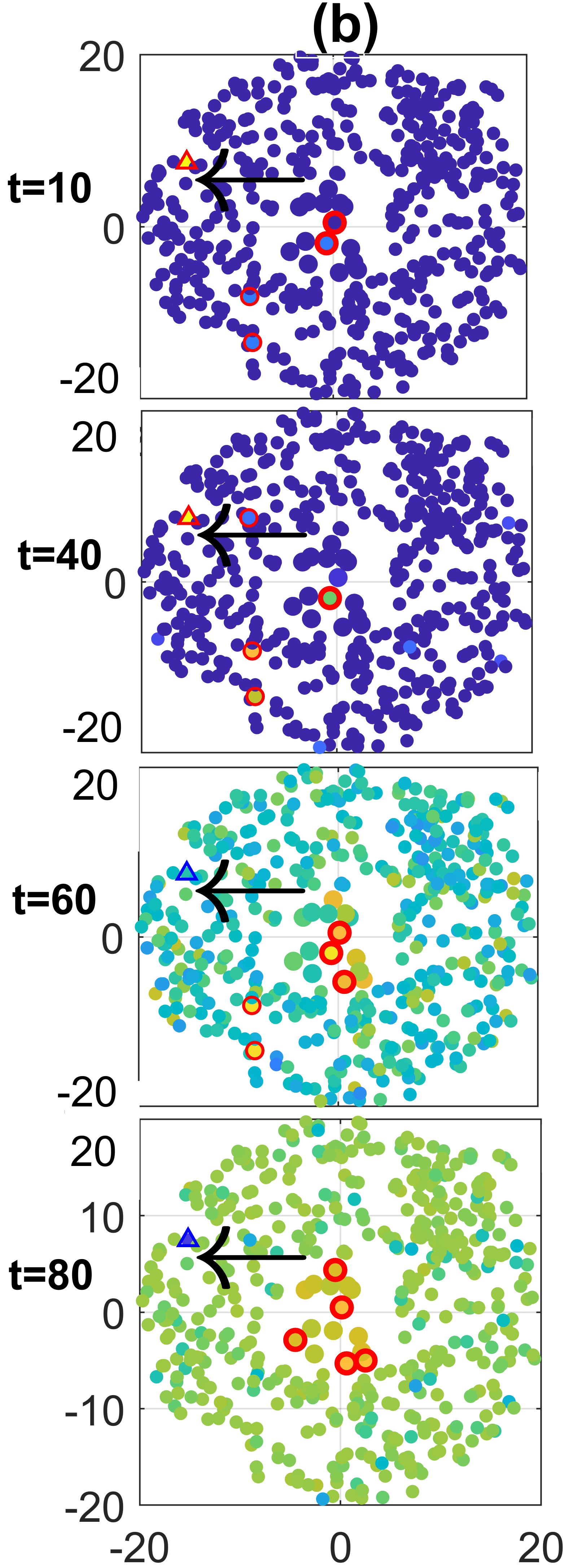} 
    \includegraphics[width=0.39\linewidth]{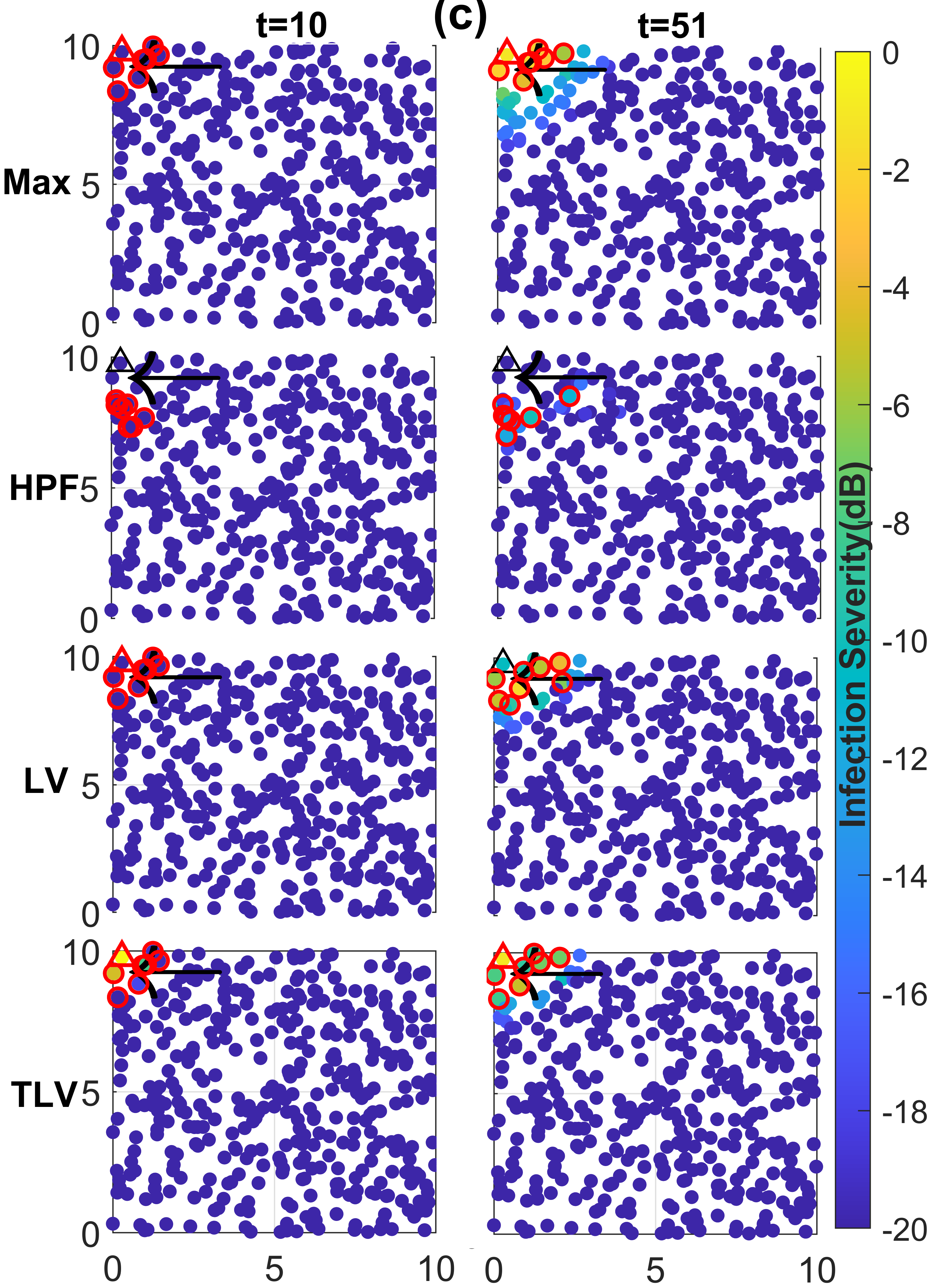}
    \caption{Evolution of infection from the infected source node of single perturbation for $\kappa=0.1$ using TLV for \textbf{(a)} Random graph and \textbf{(b)} Scale-free graph. \textbf{(c)} Identification of single and double perturbation of random graph at infected node for $\kappa=0.1$. The first and second perturbations occur at $t = 10$ and $t = 51$ respectively. Influential nodes are highlighted with red circles. Arrow is pointing towards the infected source node (marked with a triangle).}
    \label{fig:SingleDoublePertyrb}
\end{figure*}

The data $\overline{\mathbf{X}}_{var}$ obtained from Algorithm \ref{Alg:Online} is normalized to lie in the range $[0 \; 1]$. This is done using the maximum value within the window period. The scaled values are converted into decibels (dB) to aid in visualization. Negative values often occur in TLV due to directional temporal information. To avoid complex numbers arising from log-transformation of negative values, we exclude temporal variation component for nodes with negative TLV values and include only the local variation component for these nodes when plotting TLV. This is only for plotting purpose and does not affect the TLV-based identification of influential nodes. Identified influential nodes are highlighted with red circles as demonstrated in Figure \ref{fig:SingleDoublePertyrb}. The infected source node is represented as a triangle with a black arrow pointing towards it.


\subsection{Analysis of infection spread}
\label{sec:analysis_spread}
\subsubsection{Single perturbation} 
In this experiment we investigate how the disease spreads from a single infected node to the entire network. We initiate an infection at $t=1$ on a single lower degree node which propagates across the network over time ($1000$ time-steps). The parameters of SIR dynamics are: coupling coefficient $\kappa=0.1$, transmission rate $\beta_i = \beta =0.3$, and recovery rate $\gamma_i=\gamma=0.1$. Influential nodes are identified using algorithm \ref{Alg:Online} with temporal local variation ($\alpha=0.5$) and window length $r = 10$.

The disease progression and identified influential nodes at various times is shown in Figure \ref{fig:SingleDoublePertyrb}a and \ref{fig:SingleDoublePertyrb}b for random distance-based graph ($p=4\%$) and scale free graph ($p=1\%$) respectively. The node color indicates the value of TLV (normalized) for that node. In both cases, the source nodes are identified as influential node by $t=10$. The distinct progression of the disease for the two networks is clearly visible. For random distance-based graph, the spread is gradual. For the scale free graph, once the central hub nodes are infected, it rapidly spreads throughout the entire network owing to their dense connections.

\subsubsection{Double perturbation}
After the initial infection for random graph in Figure \ref{fig:graphs}a at $t=10$, at $t=50$ the infection rate $\beta$ of the same node is increased from $0.3$ to $0.8$. The results are shown in Figure \ref{fig:SingleDoublePertyrb}c with $\kappa=0.1$. We show results for TLV ($\alpha=0.7$), LV, HPF ($25$\% of high frequencies are used), and Max processing. Max processing consistently identifies the infected source node as influential. HPF is unable to identify the infected source node at both time instants. LV identified the infected source node at $t=10$, however, it failed to identify the same node at $t=51$. TLV identified the source node both the times. Superior performance of TLV is likely due to inclusion of temporal changes while calculating signal variation.

\subsection{Epidemic control}
\label{sec:EpidemicControl}
We perform simulation for epidemic control by isolating influential regions. Specifically, we simulate multiple infected nodes at multiple time steps with varying transmission rate (Section \ref{Sec:Epidemic data generation}). We apply Algorithm \ref{Alg:Online} with window of size $r=10$ to identify influential nodes. Various methods (line $3$) are used to identify influential regions - data informed methods include Max, HPF, LV, TLV, and graph topology based methods include betweenness centrality (BC) and closeness centrality (CC). 

\begin{figure}[t]
    \centering
    \includegraphics[width=0.8\linewidth]{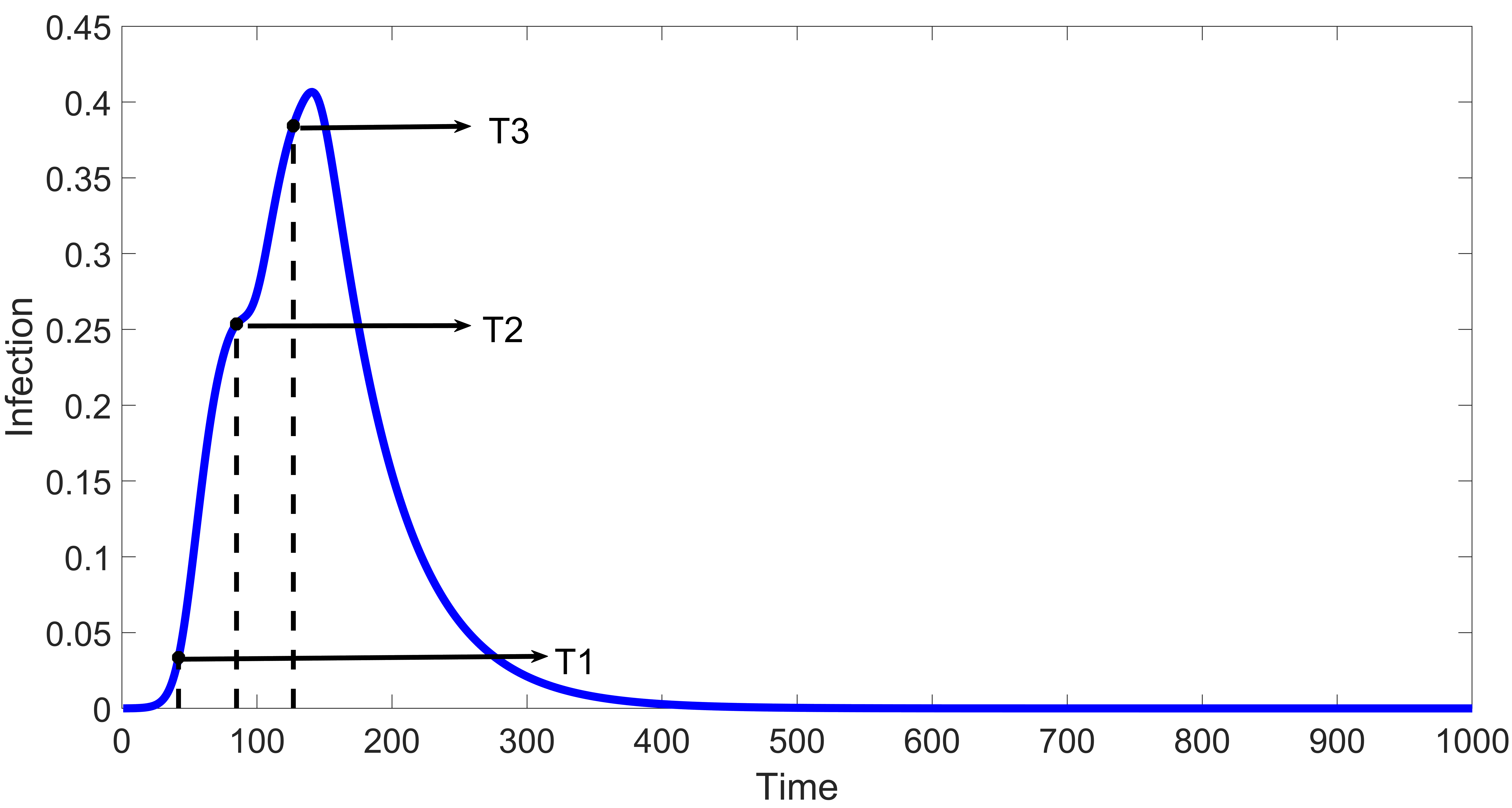} 
    \caption{Evolution of total infection across all nodes  over time}
    \label{fig:Data_timesteps}
\end{figure}

The dataset $X$ is generated using a random distance-based graph with $N = 300$ nodes and a coupling coefficient $k = 0.1$ over $1000$ time steps. The total infection is calculated by summing the infections of all nodes at each time step, providing a time series of the total infection. The peak infection time, denoted as $T$, is identified and used for further processing. Randomly chosen five nodes are infected at $t = 1$, initiating the infection. This is repeated three more times at $T_1$, $T_2$, and $T_3$ by randomly choosing five nodes and increasing their infection rate to $0.6$. We choose $T_1 = 0.3 T$, $T_2 = 0.6 T$, and $T_3=0.9 T$ where $T$ is the time at which peak of total
infection propagation occurs before applying control. Figure \ref{fig:Data_timesteps} illustrates the total infection summed across all the nodes when no intervention is applied.

\begin{figure}[t]
    \centering
    \includegraphics[width=0.9\linewidth]{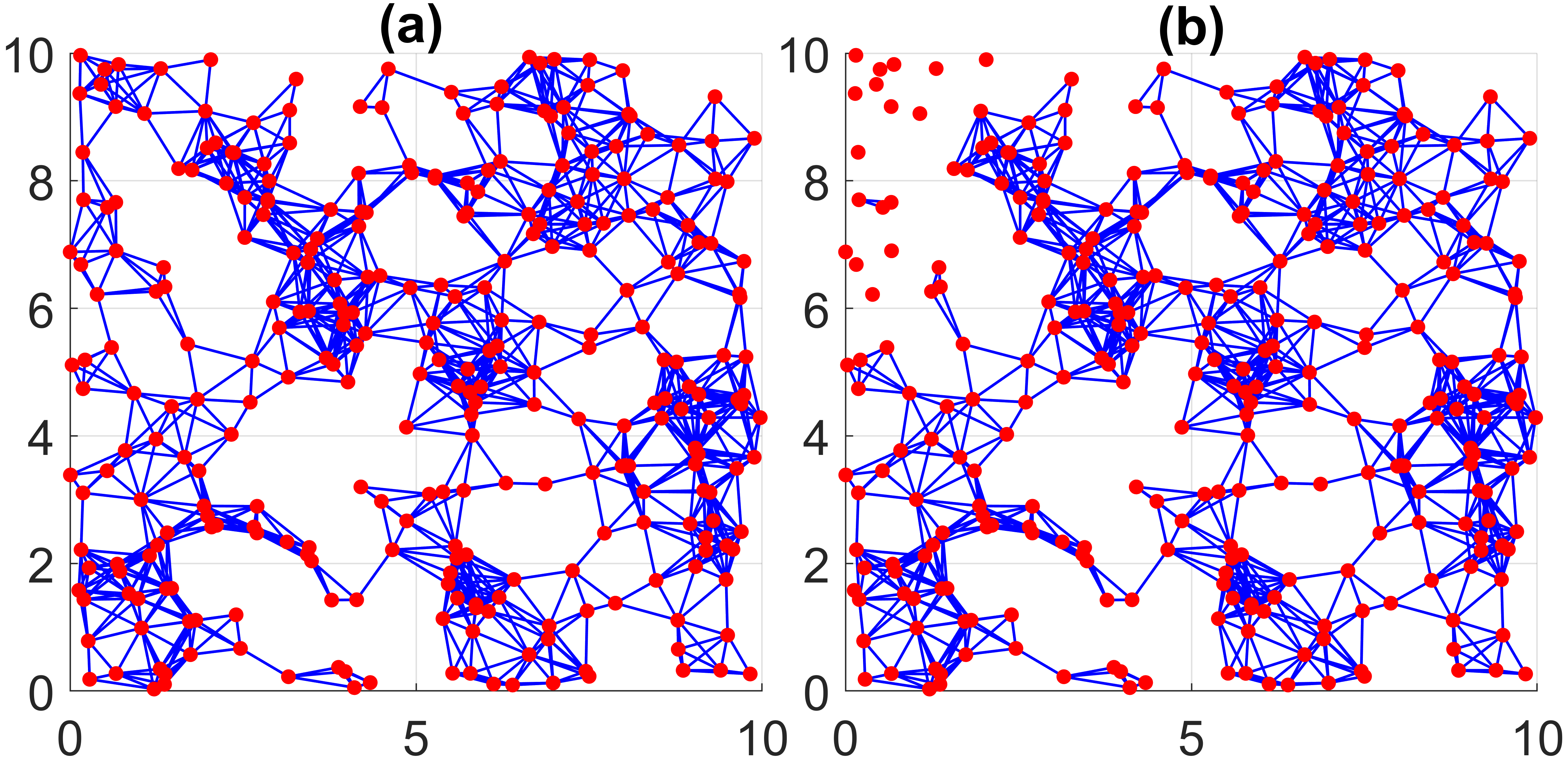} 
    \caption{Comparison of a) Original Graph $\mathcal{G}$ and b) Control Graph $\mathcal{G}_1$ at $T_1$ using TLV.}
    \label{fig:ControlGraph}
\end{figure}

\begin{figure}[t]
    \centering
    \includegraphics[width=0.9\linewidth]{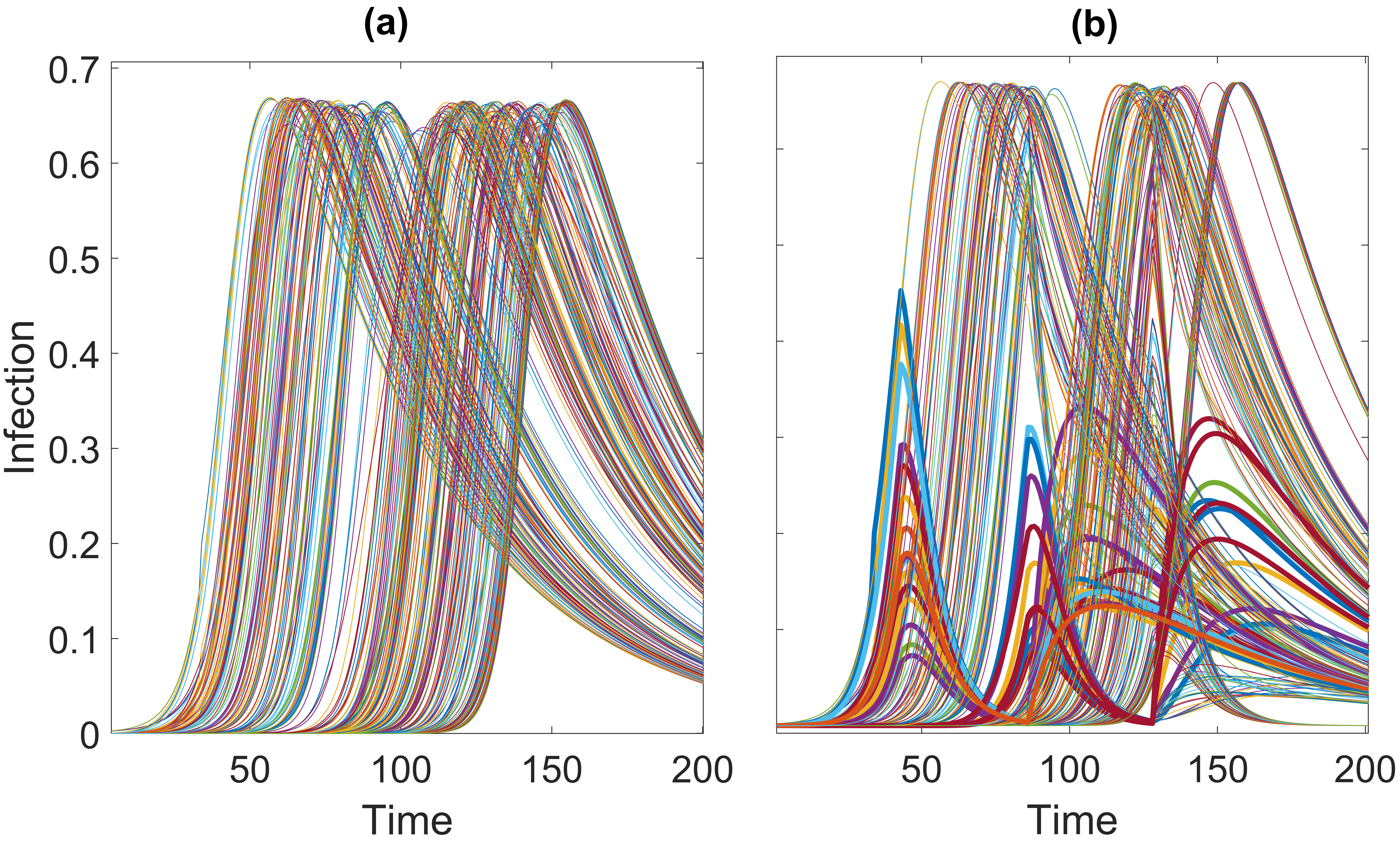}
    \caption{Infection and control measures: (a) Infection without any control measure (b) Infection with epidemic control on influential nodes identified using TLV.}
    \label{fig:data_before_after_control}
\end{figure}

Control is applied at multiple stages as presented in Algorithm \ref{Alg:ControlAlgorithm for n stages}, producing $\mathbf{X}_c$ -- infection levels of all nodes at all times after implementing the control measures. The $n^{th}$ control intervention is applied at time $T_n + t_c$. Specifically, in the $n^{th}$ intervention, we disconnect links of $p_n\%$ of nodes (influential nodes identified using TLV). Influential nodes identified at consecutive control time steps are not likely to repeat because disconnected nodes have zero local variation. This results in a lower TLV value influenced only by the temporal component, which is often expected to be negative. We set $n = \{1,2,3\}$, $p_1=5\%$, $p_2=10\%$, $p_3=15\%$, and $t_c=10$. Figure \ref{fig:ControlGraph} shows the original graph and the control graph at time $T_1$ after intervention using TLV. For all the $300$ nodes, the infection evolution is plotted in Figure \ref{fig:data_before_after_control}. Figure \ref{fig:data_before_after_control}(a) shows the infection without any control ($\mathbf{X}_c$), while Figure \ref{fig:data_before_after_control}(b) shows the $\mathbf{X}_c$, infection when control measures are implemented on influential nodes identified using TLV ($\alpha=0.6$). The reduction in infection is observed in Figure \ref{fig:data_before_after_control}(b) (marked with thicker lines in the figure), at time steps $T_1$, $T_2$, $T_3$ especially among influential nodes whose links have been strategically removed.
 
\begin{algorithm}
\caption{Influential Identification and Epidemic Control for $n$ stages}
\begin{algorithmic}[1]
    \State \textbf{Input:} $X$, $\mathcal{G}$, \{$T_1$, $T_2$...$T_n$\}; \{$p_1$,$p_2$...,$p_n$\} (n \text{is the number of interventions})
    \State \textbf{Output:} $X_c$ for different methods

    \For{$i \gets 1$ \textbf{to} n }
        \State $Ti \gets Ti + t_c$

        \If{$i = 1$}
        \State Apply Algorithm \ref{Alg:ControlAlgorithm} on $\mathcal{G}$ at $T_1$ and generate control graph $\mathcal{G}_1$ by disconnecting $p_1\%$ of nodes
    
        \Else
        \State Apply Algorithm \ref{Alg:ControlAlgorithm} on $\mathcal{G}_{i-1}$ at $T_i$ and generate control graph $\mathcal{G}i$ by      
             disconnecting $p_i\%$ of nodes from 
            $\mathcal{G}$ to obtain $\mathcal{G}_i$ and
            use it for disease propagation

    \EndIf
    \EndFor
    \State \textbf{Calculate $\mathbf{X}_c$  for different methods using $\mathcal{G}_n$ (final control graph)}
\end{algorithmic}
\label{Alg:ControlAlgorithm for n stages}
\end{algorithm}

\subsection{Epidemic control -- analysis of parameter $\alpha$}
\label{sec:AlphaAnalysis}

\begin{figure*}
    \centering
    \subfloat[Distribution of best $\alpha$]{
        \includegraphics[width=0.43\linewidth]{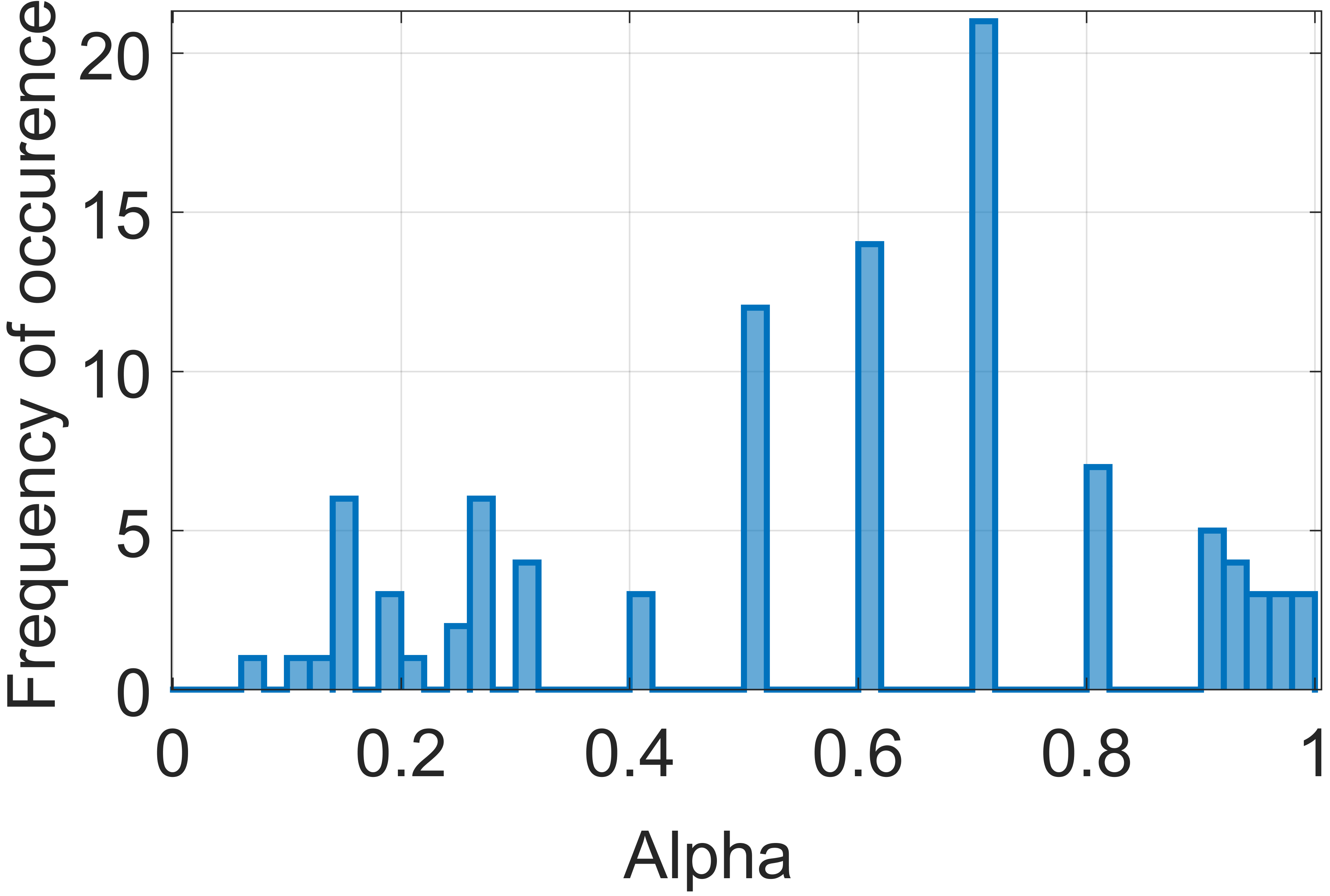}
    }
    \hspace{0.02\linewidth} 
    \subfloat[Cumulative infection]{
        \includegraphics[width=0.46\linewidth]{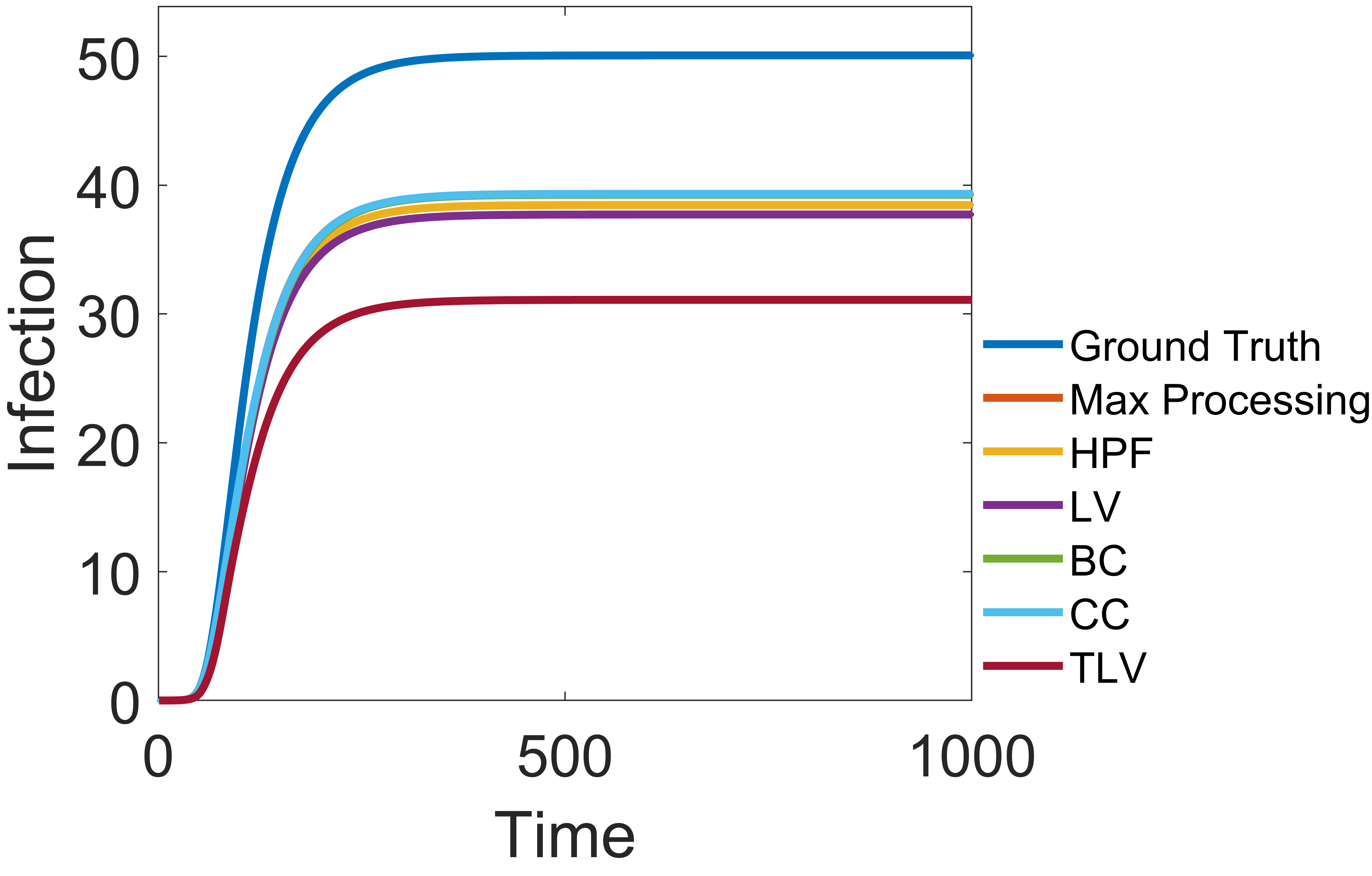}
    }
    \caption{Distance-based graph: (a) Histogram of best $\alpha$ values over 100 MC trials and (b) Monte Carlo average of cumulative (across network) infection as time progresses.}
    \label{fig:Hist_Cumulative_Random}
\end{figure*}

To investigate the effect of parameter $\alpha$ and to identify its optimal range, we perform Monte Carlo trials. The epidemic control experiment (Section \ref{sec:EpidemicControl}) on the distance-based graph is repeated for $100$ trials with the set of infected nodes randomly chosen in each trial. The experiment was repeated for $\alpha$ values in the set $\{ 0:0.02:0.3, 0.4:0.1:0.9, 0.9:0.02:1 \}$. Finer granularity is selected at extreme parameter values as higher sensitivity was noted in this regime. For each $\alpha$, the cumulative (across nodes) infection is computed for each trial. The ``best $\alpha$'' is identified as the $\alpha$ value that resulted in the least cumulative infection at the end of the simulation. Thus, we obtain a list of $100$ best-performing $\alpha$ values. Figure \ref{fig:Hist_Cumulative_Random}a displays the histogram of best $\alpha$ values. It suggests that $\alpha$ values greater or equal to $0.6$ should be preferred. 

The average (across MC trials) of the cumulative infection for all trials with best $\alpha \geq 0.6$ is shown in Figure \ref{fig:Hist_Cumulative_Random}b. For TLV, the simulation corresponding to the ``best $\alpha$'' is used in each trial. The ``Ground truth''refers to the average cumulative infection in absence of any control strategy. The results demonstrate that TLV consistently outperforms other methods. Specifically, the methods ranked from highest to lowest average cumulative infection are -- CC, Max, BC, HPF, LV, and TLV. This clearly demonstrates that TLV outperforms other metrics which only use the graph topology while ignoring the data (CC and BC) as well as fully data-based method which ignores the graph topology (Max).

\begin{figure*}
    \centering
    \subfloat[Distribution of best $\alpha$]{
        \includegraphics[width=0.40\linewidth]{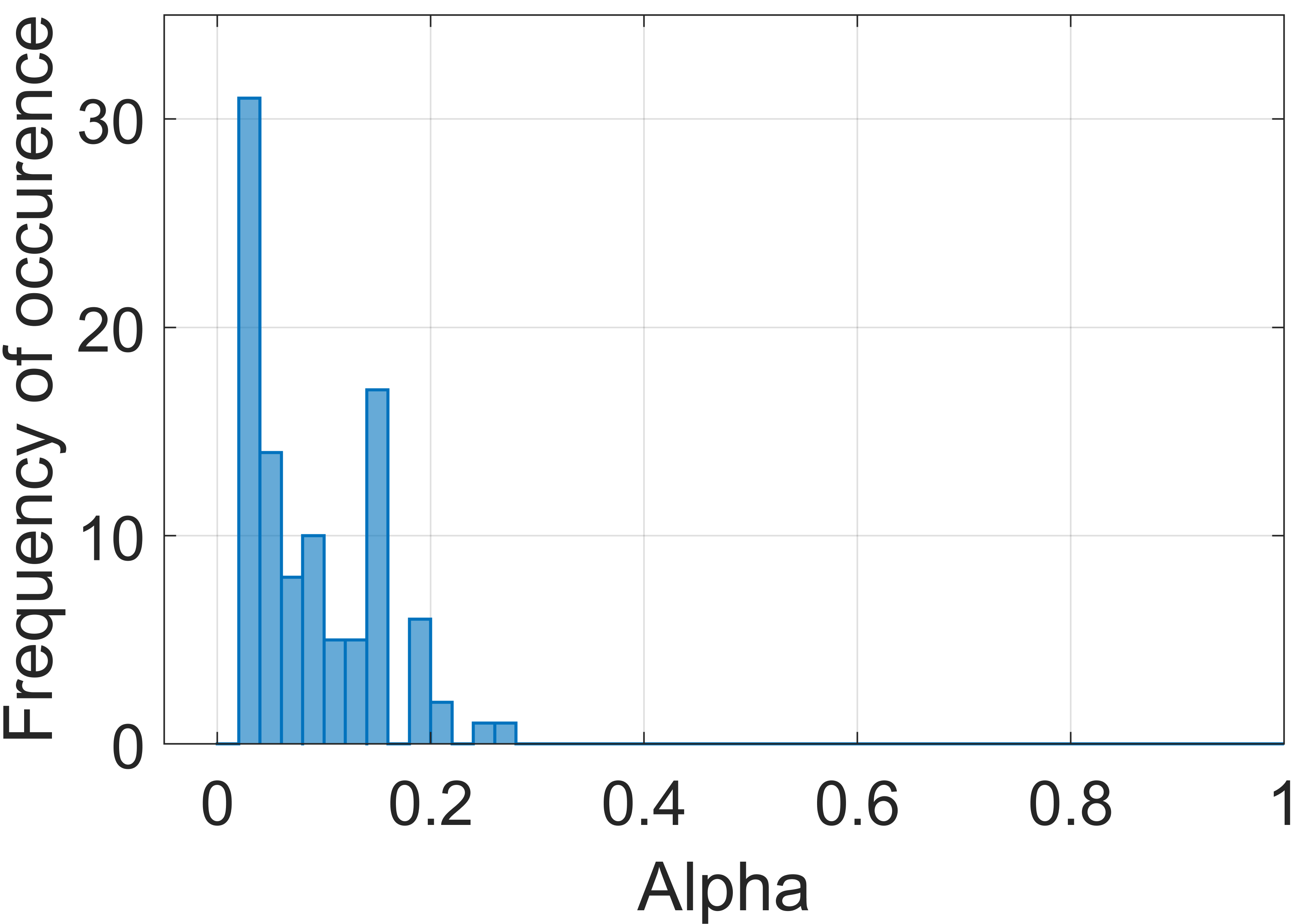}
    }
    \hspace{0.02\linewidth} 
    \subfloat[Cumulative infection]{
        \includegraphics[width=0.49\linewidth]{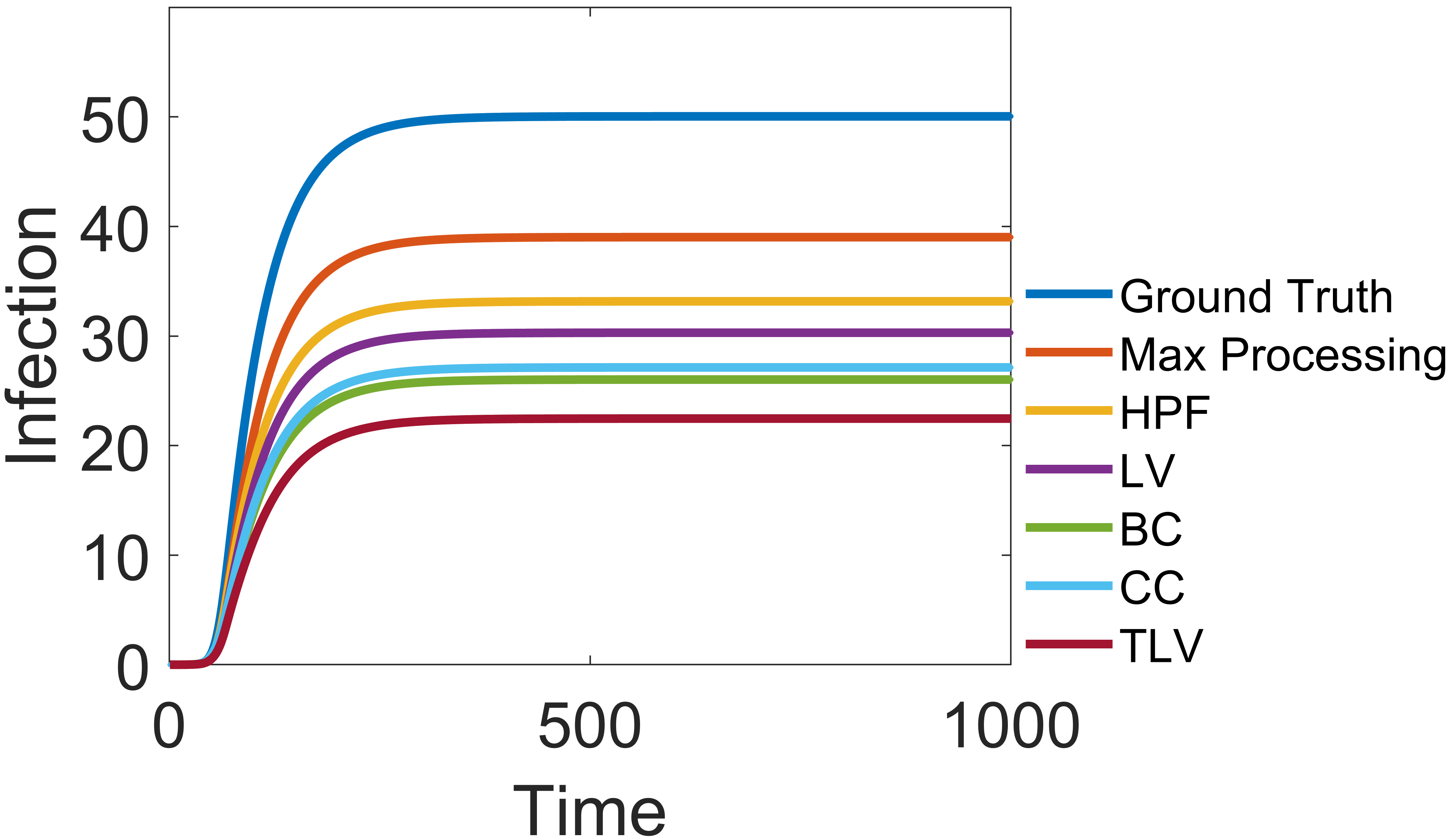}
    }
    \caption{Scale-free graph: (a) Histogram of best $\alpha$ values over 100 MC trials and (b) Monte Carlo average of cumulative (across network) infection as time progresses.}
    \label{fig:Control measures for scalefree}
\end{figure*}

We perform a similar analysis using a scale-free graph with $300$ nodes. Data generation and processing details remain the same. The histogram of optimal $\alpha$ values is shown in Figure \ref{fig:Control measures for scalefree}a. For high coupling coefficient $(\kappa = 0.1)$, histogram is concentrated at small $\alpha$ values. The average cumulative infection is shown in Figure \ref{fig:Control measures for scalefree}b.  


\subsection{Comparison with SGWT analysis} 
\label{Sec:BatchProcessing}

Spectral graph wavelet theory (SGWT) has been applied for the influential node identification \cite{geng2022analysis}. We perform similar approach as given in \cite{geng2022analysis}. Epidemic data is simulated for $350$ time steps (days) on a $400$ node random distance-based binary graph. Network SIR dynamics are simulated with $\beta=0.3$ and $\gamma=0.03$. The infection is initiated from a randomly selected node. We obtain a sliding two-week cumulative infection data with slide length of one week. Let this biweekly infection data be denoted $\overline{\mathbf{X}}$.

\subsubsection{Influential nodes} 
Influential nodes are identified in \cite{geng2022analysis} using SGWT analysis on the spatio-temporal graph constructed using the graph strong product. This analysis determined a total of $S = 1176$ influential spatio-temporal nodes for the entire period. We compare this with influential nodes obtained using TLV in Algorithm \ref{Alg:Online}. We first compute the TLV of the input signal to obtain $\overline{\mathbf{X}}_{var}$. Subsequently, for each day, the top $k$ nodes with highest TLV values are identified as influential nodes where $k$ was determined as $S/\mathcal{T} = 1176/49 = 24$. For max processing we set $\overline{\mathbf{X}}_{var}=\overline{\mathbf{X}}$.

\begin{figure*}[t]
    \centering
    \includegraphics[width=1\linewidth]{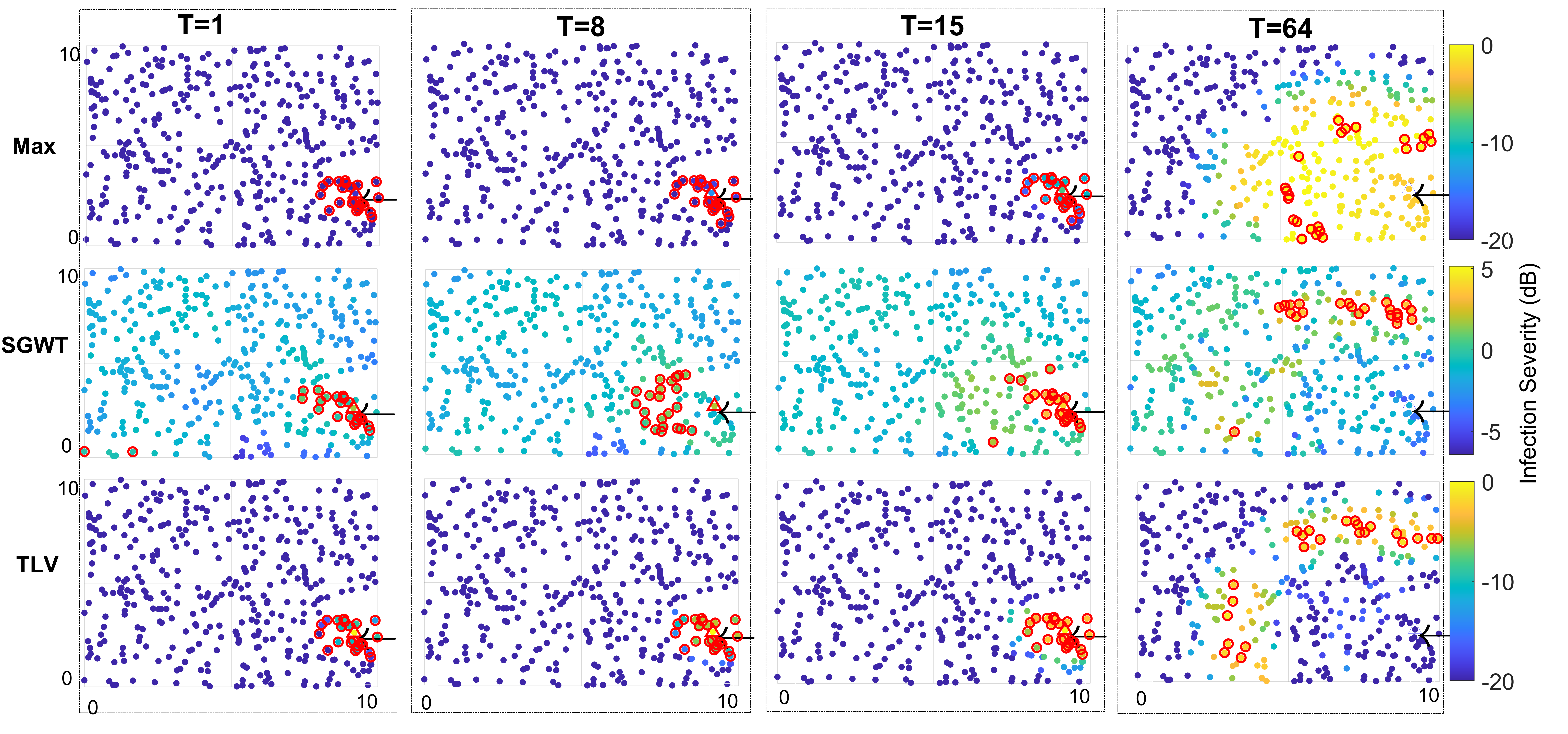}
    \caption{Comparison of Max, SGWT and TLV in influential node identification for a distance-based graph at different time steps. Influential nodes are highlighted with red circles.}
    \label{fig:batch_topK_threemethods}
\end{figure*}

\subsubsection{Observations}

Figure \ref{fig:batch_topK_threemethods} visualizes the influential nodes identified using Max, SGWT, and TLV. The columns demonstrate epidemic spread patterns at various time steps.
In the initial period ($T = 1$ to $14$, i.e., first two weeks), all methods successfully identify the infected source node and neighboring nodes as influential. Additionally, SGWT identifies a few nodes (bottom left part of the panel) located farther from the source. These faraway nodes have lower infection rates and are not expected to exert significant influence.

As the value of $T$ increases from left to right in the Figure \ref{fig:batch_topK_threemethods}, we notice a similarity between the nodes identified using Max and TLV. However, the behavior exhibited by the SGWT method reveals a different pattern. Specifically, as the time progresses, the influential nodes identified by SGWT seems to drift away from the source node and then come back. This reversal of the wave front pattern is unexpected and deviates from the anticipated spread dynamics.



\subsection{H1N1 data}
The H1N1 influenza virus \cite{smith2009origins}, commonly known as swine flu, emerged in Mexico in $2009$ and is of global health concern. The world health organization declared it a pandemic as the virus quickly traversed borders, affecting populations worldwide. We study a hybrid dataset which integrates real airport network with simulated H1N1 dynamics. Our task is to identify influential regions for epidemic control.

\begin{figure}[t]
    \centering
    \includegraphics[width=0.9\linewidth]{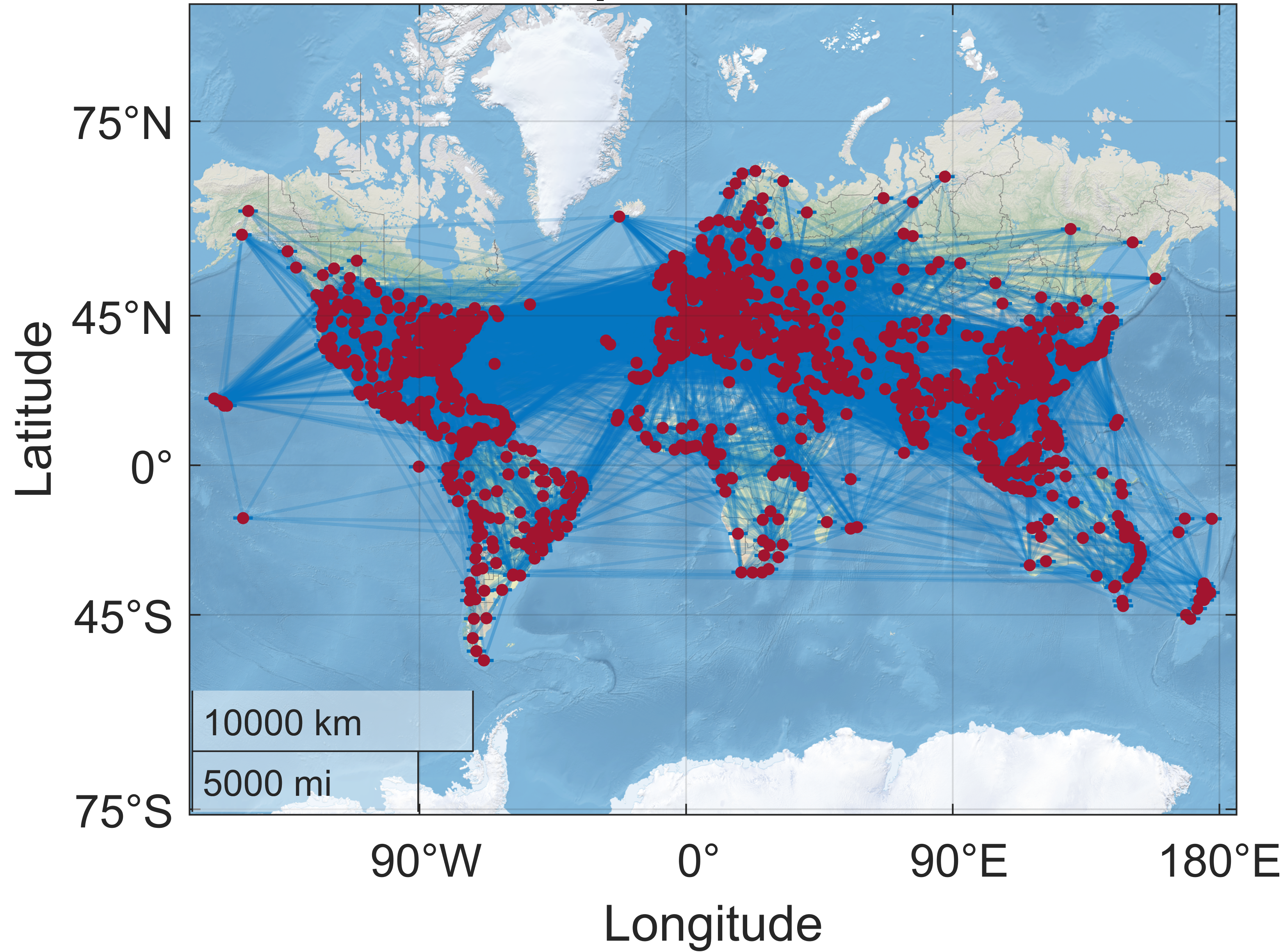}
    \caption{Airport Network with $1292$ nodes (airports) and $?$ connections.}
    \label{fig:Airport Network}
\end{figure}

\begin{figure*}[t]
   \centering
    \includegraphics[width=1\linewidth]{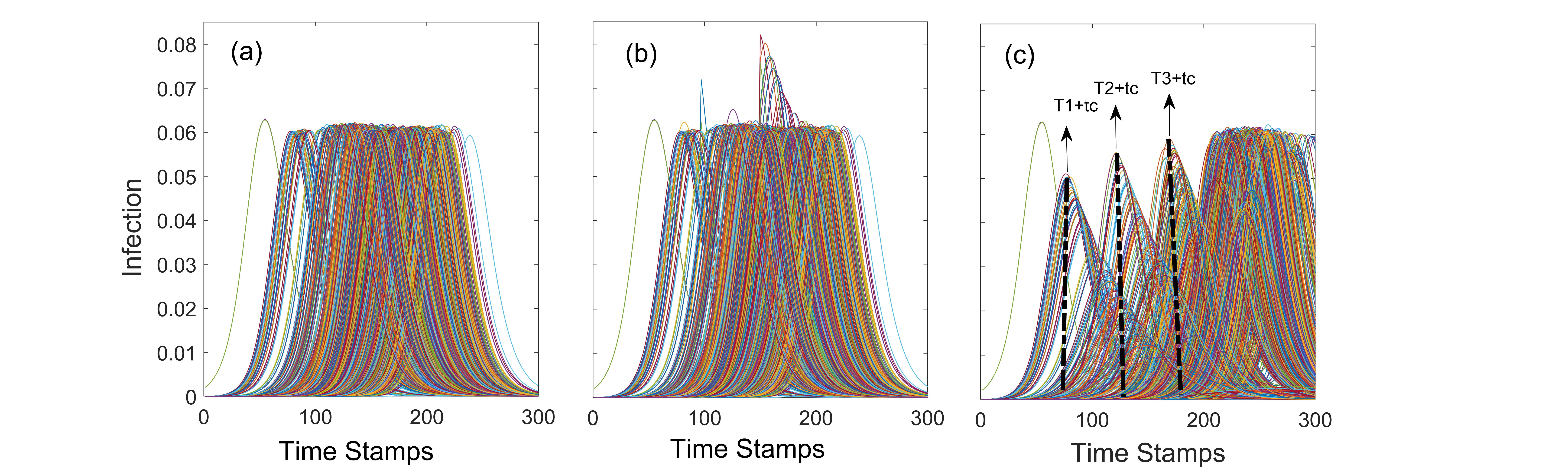} 
   \caption{Comparison of a) Original infection on graph $\mathcal{G}$ and b) Infection after increased transmission rate at selected nodes at $T_1$, $T_2$, $T_3$ c) Infection after applying control measures at times $\{T_1, T_2, T_3\} + t_c$}
    \label{fig:Data_H1N1}
\end{figure*}

\begin{figure*}[t]
    \centering
    \includegraphics[width=0.9\linewidth]{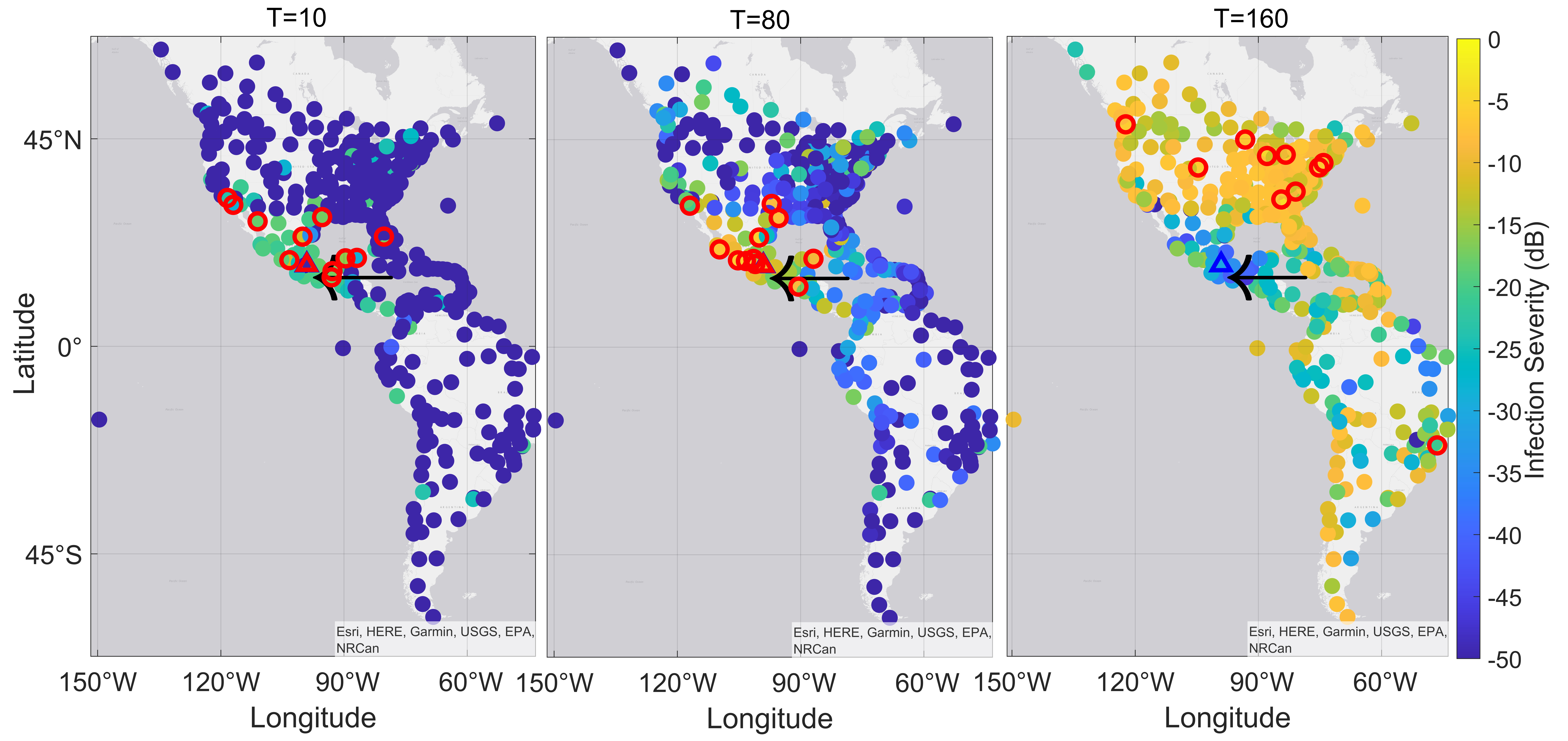} 
    \caption{Identified influential nodes and infection spread over time steps using TLV}
    \label{fig:H1N1_spread}
\end{figure*}

\begin{figure}[t]
    \centering
    \includegraphics[width=0.9\linewidth]{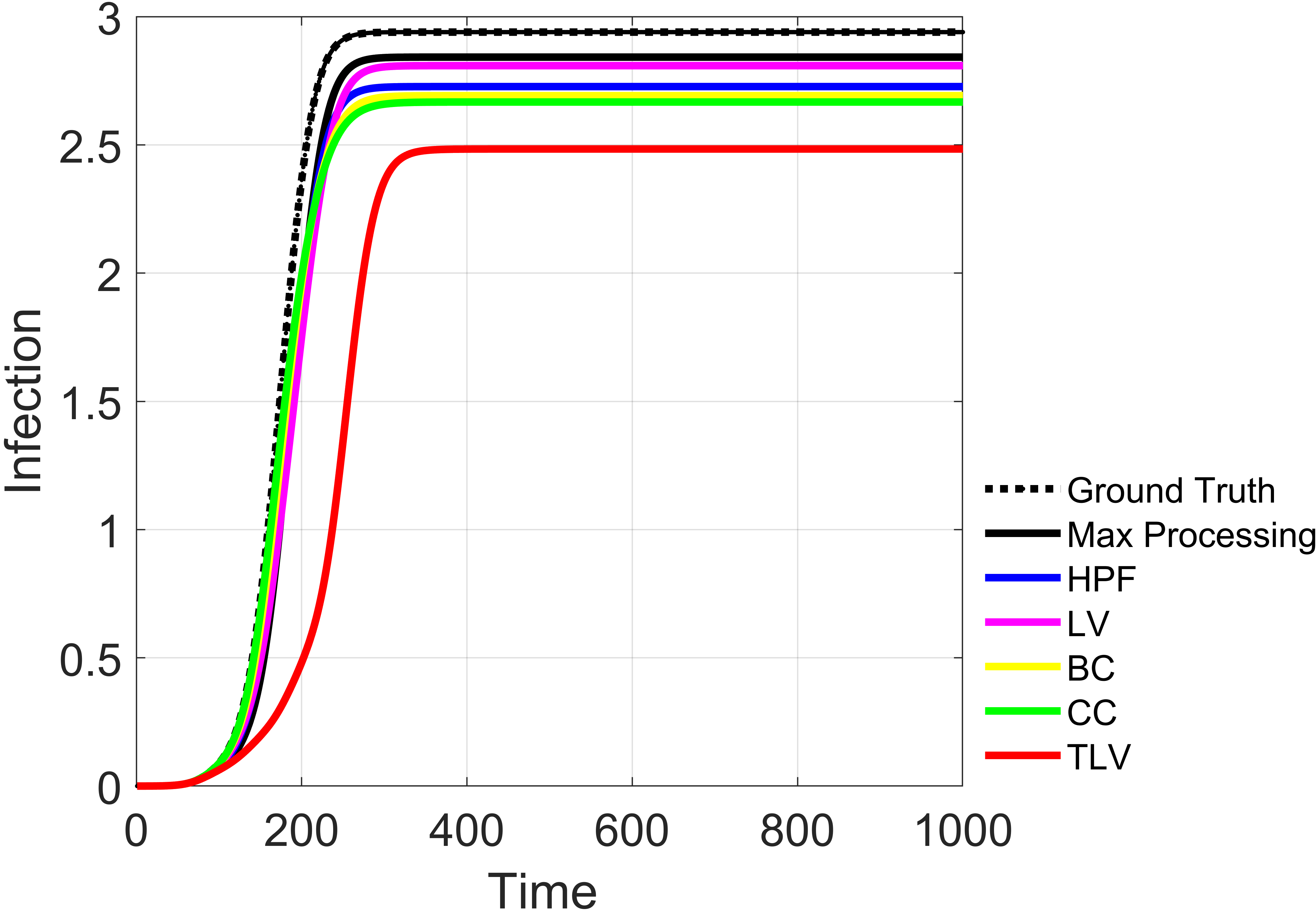}
    \caption{Comparison of cumulative infection for H1N1 data using different methods }
    \label{fig:Cumulative_H1N1}
\end{figure}

Work in \cite{brockmann2013hidden} investigated the dynamics of H1N1 contagion phenomena using real airport network, exploring the hidden geometric patterns and the parameters that influence the spread of disease. Leveraging the findings on disease parameters from \cite{brockmann2013hidden}, we generate H1N1 data using the airport network of $1292$ locations, passenger data from $2009$, coupling constant $\kappa=0.0028$, and initial infection source in Mexico. Figure \ref{fig:Airport Network} illustrate the airport network and Figure \ref{fig:Data_H1N1}a shows the simulated H1N1 infection at all the nodes. The airport network exhibits a scale-free topology, evident in the presence of highly connected airport hubs. The infection data is analyzed and the temporal local variation is plotted in Figure \ref{fig:H1N1_spread} at various time points. It can be seen that Mexico is identified as an influential node at $T=10$ in Figure \ref{fig:H1N1_spread}a. Subsequent subplots, generated with Algorithm \ref{Alg:Online} $(r = 10)$ using TLV, depict the pandemic's local propagation and top $p = 1\%$ influential nodes. 
We now increase the infection transmission rate to $0.8$ for $2\%$ of the randomly selected nodes at times $T_1$, $T_2$, and $T_3$. The resulting infection data is shown in Figure \ref{fig:Data_H1N1}b. The elevation of the transmission rate serves the purpose of introducing a realistic representation of super-spreader events into the H1N1 dynamics, mirroring occurrences in real-world scenarios. We now implement epidemic control at times $\{T_1, T_2, T_3\} + t_c$ as outlined in section \ref{sec:EpidemicControl} using Algorithm \ref{Alg:ControlAlgorithm for n stages}. Specifically, links from $5\%$, $10\%$, and $15\%$ of nodes are removed $t_c$ time steps after $T_1$, $T_2$, and $T_3$ respectively. Our findings reveal that, for TLV, $\alpha = 0.9$ is most effective in infection reduction and the resulting infection levels are shown in Figure \ref{fig:Data_H1N1}c. This is consistent with our observations in Section \ref{sec:AlphaAnalysis} that, for scale-free networks, higher $\alpha$ values lead to highest infection reduction when $\kappa$ is small. The cumulative total infection is shown in Figure \ref{fig:Cumulative_H1N1}.

\section{Analysis of Indian COVID-19 Data}
\label{sec:realdata}

We analyzed confirmed (i.e. infected) COVID-19 cases across 630 districts in India, sourcing data from \cite{Covid_19_india}.
To ensure fair comparison of districts of different sizes, we normalize confirmed cases by the district population \cite{wikipedia:List_of_districts_in_India}. This normalized COVID-19 confirmed cases in each district forms the graph signal $\mathbf{X}$. A network $\mathcal{G}$ with districts as nodes is built based on Euclidean distance \eqref{Weights}, where nodes within a defined threshold ($\vartheta$) are connected, and closer connections are assigned higher weight. This approach is motivated by the predominant use of road transportation in India, where disease transmission is more likely to occur between nearby nodes due to population movement patterns. The distance between districts is calculated using their district headquarters, and the threshold $\vartheta$ is set as 100 km (calculated based on mean distance).

\begin{figure*}
    \centering
    \subfloat[Indian Farmers Protest]{
        \includegraphics[width=0.42\linewidth]{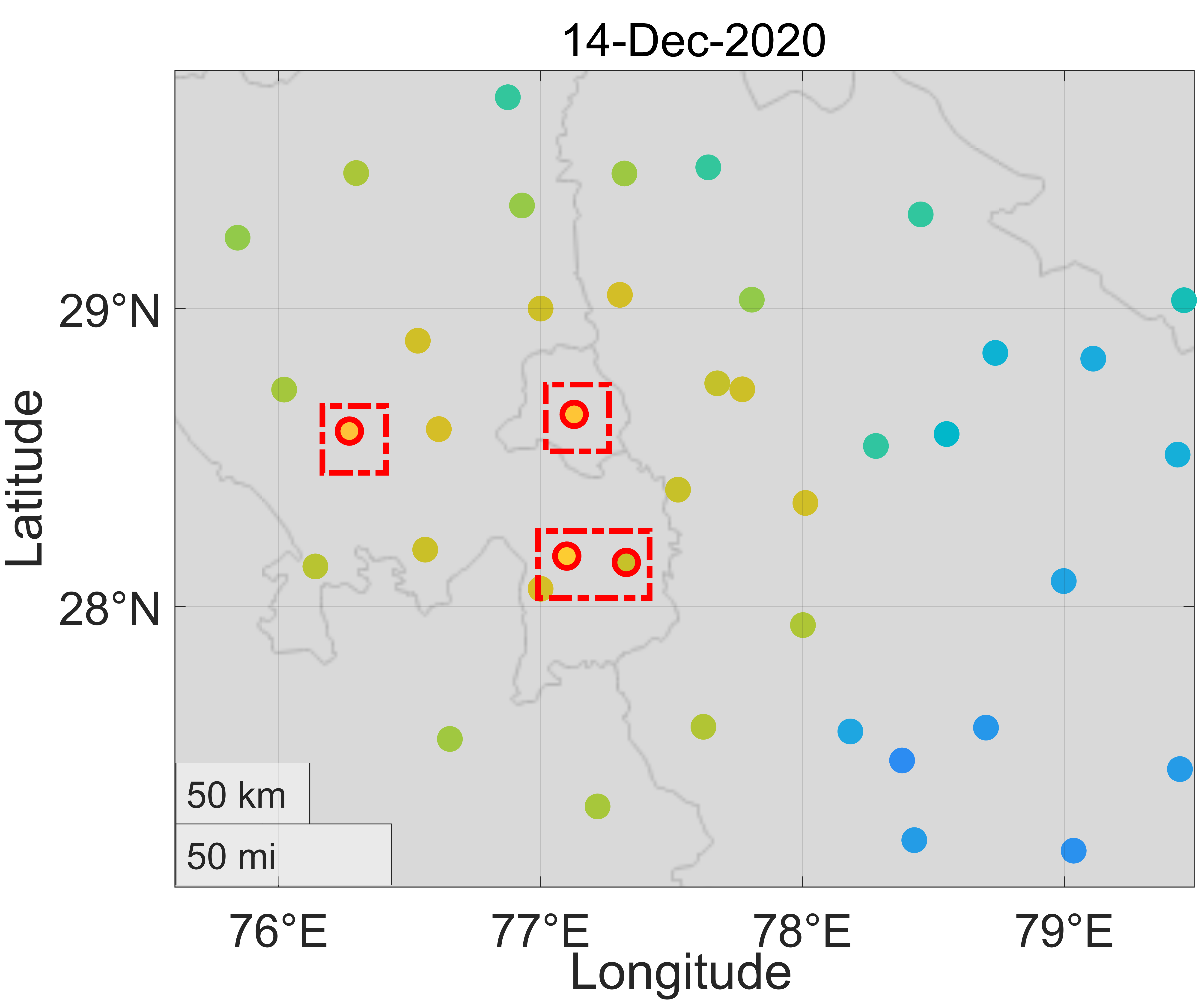}
    }
    \hfill
    \subfloat[Kumbha Mela]{
        \includegraphics[width=0.51\linewidth]{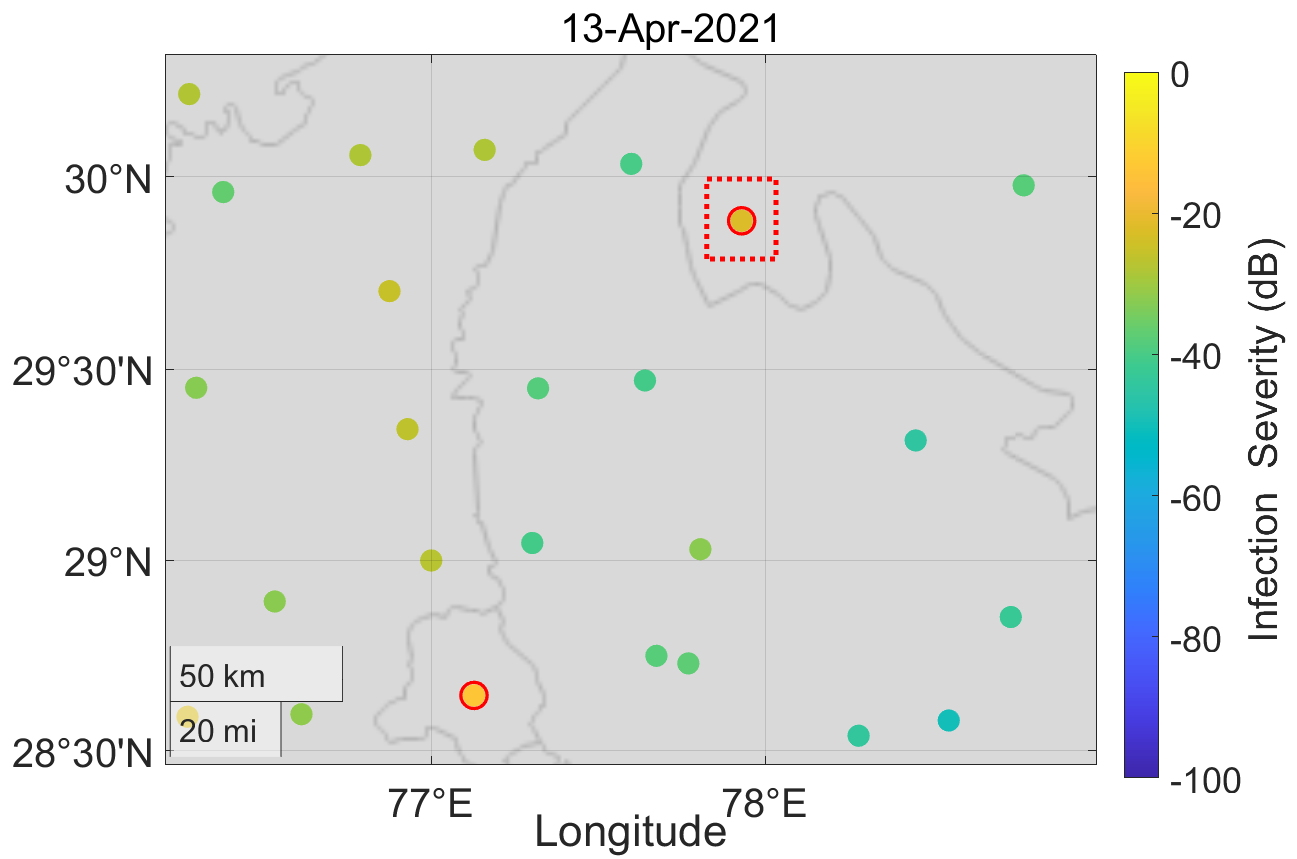}
    }
    \vskip\baselineskip
    \subfloat[Elections in Puducherry and Tamil Nadu]{
        \includegraphics[width=0.435\linewidth]{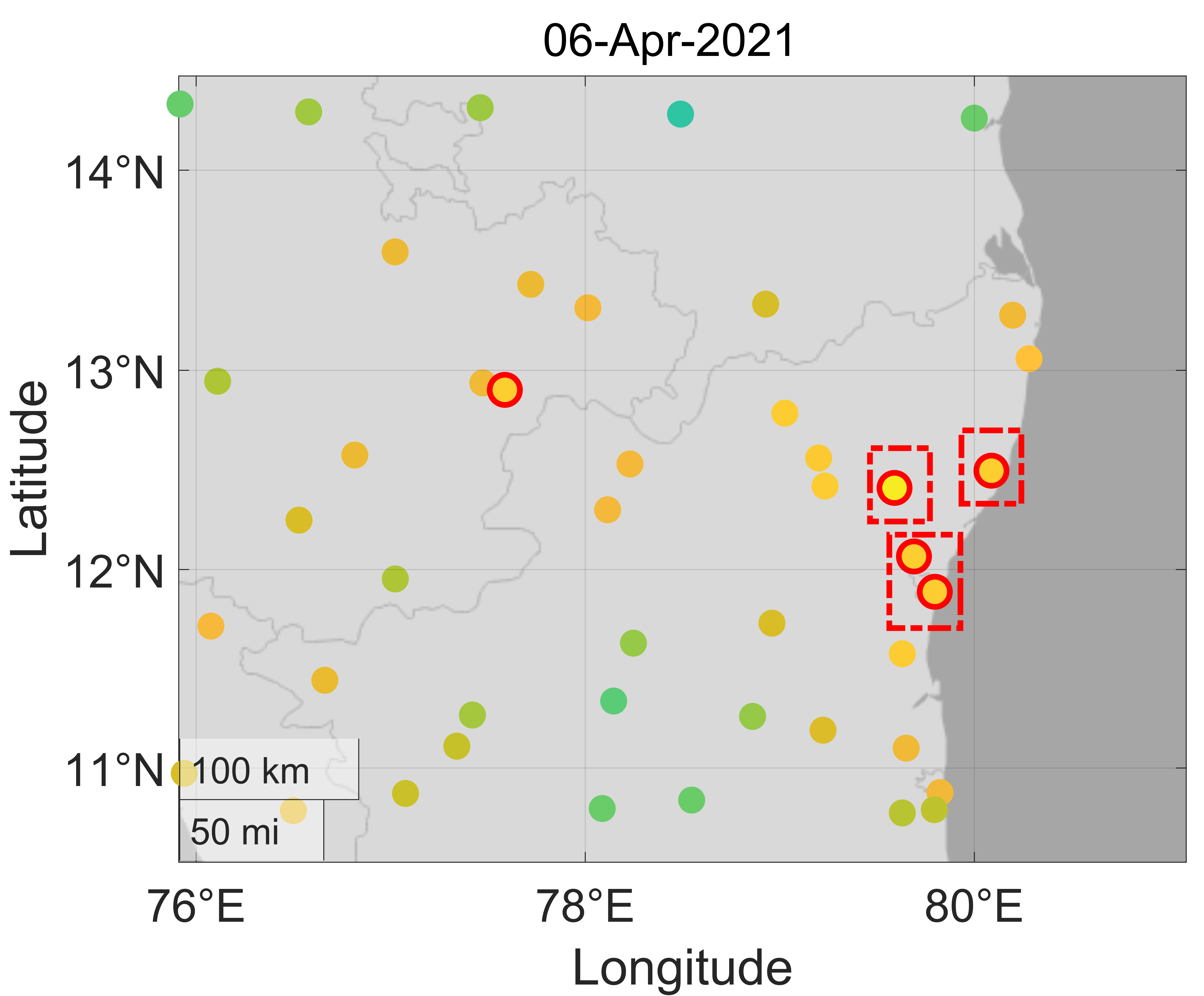}
    }
    \hfill
    \subfloat[Maharashtra as hotspot]{
        \includegraphics[width=0.515\linewidth]{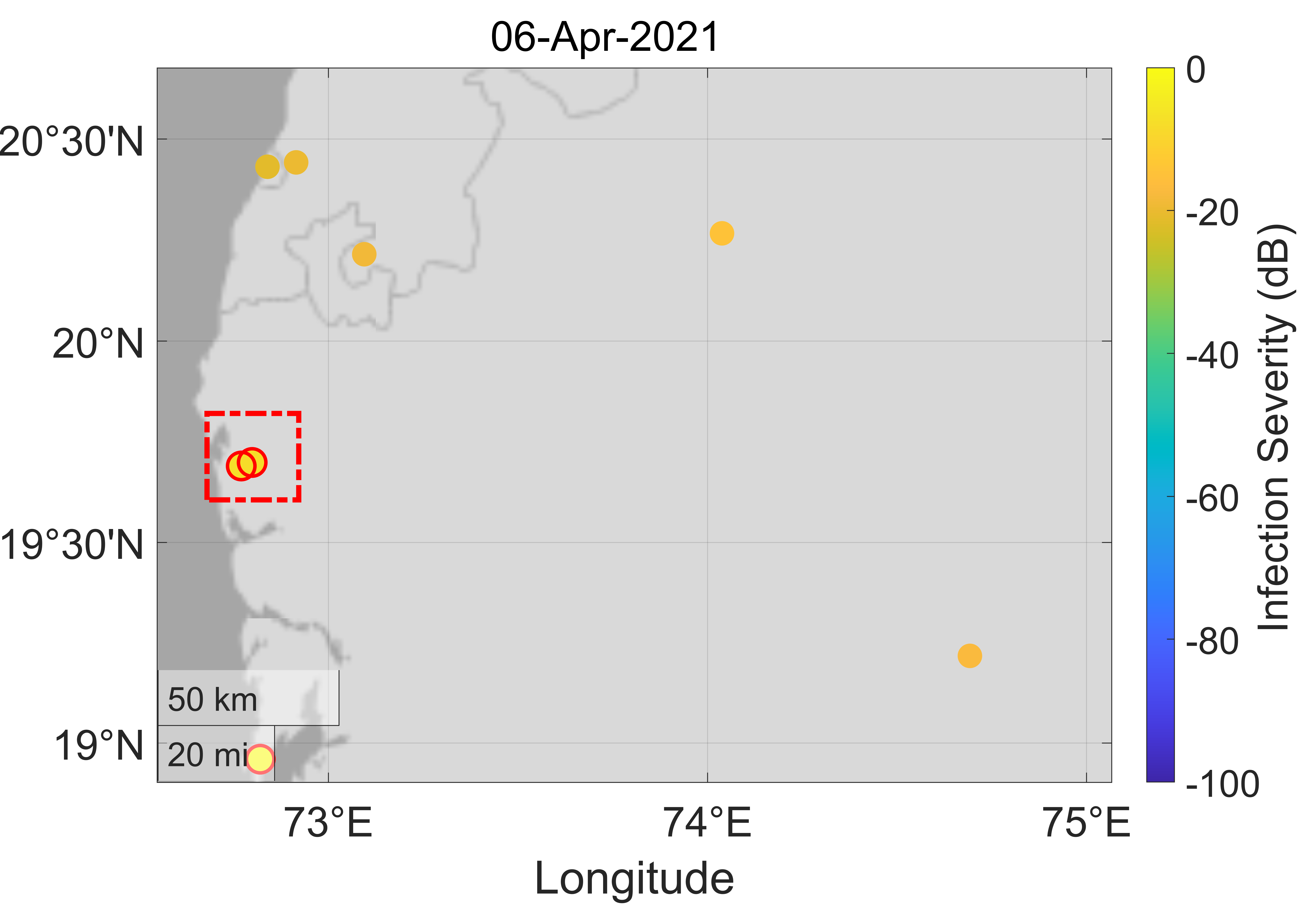}
    }
    \caption{Indian COVID-19 data analysis using TLV: Identified influential nodes are shown as red circles. The red squares correspond to regions (as identified in the news) which subsequently led to a massive spike in the number of infected cases.}
    \label{fig:SuperspreaderEvents}
\end{figure*}

Based on the analysis in Section \ref{sec:AlphaAnalysis}, the optimal value of $\alpha$ for TLV for the distance-based graph is $\alpha \geq 0.5$. Thus, for identification of influential regions (districts) in India, we considered $\alpha=0.6$. We choose $r=10$ and identify top $p\%=1\%$ nodes as influential. Our analysis reveals that some of the identified influential nodes have been attributed to a subsequent large spike (as corroborated by news articles) in COVID-19 cases resulting from large gatherings. Instances of this include:
\begin{itemize}
    \item On December $14$, $2020$, tens of thousands of Indian farmers protested on the outskirts of New Delhi, calling for a nationwide farmers' strike \cite{kumar2020significance}, as shown in Figure \ref{fig:SuperspreaderEvents}a.
    \item On January $14$, $2021$, the Kumbh Mela-$2021$ emerged as a potential super spreader event \cite{kumbhmela}, in Uttarakhand, lasting until April $29$, $2021$.  Figure \ref{fig:SuperspreaderEvents}b illustrates one of the date within the event period when Uttarakhand is identified as a key influential region.
     \item Political rallies \cite{firstpost2021} during state assembly elections in few states exacerbated COVID-19 spread. As shown in Figure \ref{fig:SuperspreaderEvents}c, Puducherry and Tamil Nadu are identified as influential region in April first week of $2021$.
     \item Maharashtra continued to be a COVID-19 hotspot throughout $2021$, experiencing several waves of rising cases and fatalities \cite{IndiaToday2021,balasubramani2024spatio,wikipedia2024,indianexpress2021}. Figure \ref{fig:SuperspreaderEvents}d llustrates Maharashtra as a influential region. Our algorithm identifies it as influential region most of the times from $2020$ to $2021$.
\end{itemize}

\section{Conclusion}

We introduced a variation measure for temporally evolving graph signals. The proposed metric, temporal local variation (TLV), accounts for signal variation in the graph domain as well as in the time domain. Desired balance between spatial (graph) and temporal variation can be chosen by appropriate choice of the parameter $\alpha$. Building on this, an algorithm is developed to rank nodes according to their influence in signal propagation. The algorithm is applied on infection data to identify influential pandemic regions. Control experiments demonstrate that the TLV is effective at identifying influential nodes which in turn leads to largest reduction in cumulative infection. We perform experiments to identify optimal range of the parameter $\alpha$ for various networks and dynamics (coupling strengths). In addition to simulated data, we validated our method using hybrid H1N1 dataset and Indian COVID-19 data.

\bibliographystyle{IEEEbib}
\bibliography{refs}

\end{document}